\documentclass[useAMS,usenatbib,usegraphicx]{mn2e}

\usepackage{subfigure}
\usepackage{url}
\usepackage{times}
\usepackage{comment}
\usepackage{color}
\usepackage{dcolumn}
\usepackage{enumitem}

\makeatletter

\renewcommand{\@thesubfigure}{(\alph{subfigure})\hskip\subfiglabelskip}
\renewcommand{\@@thesubfigure}{(\alph{subfigure})}

\newcommand\ion[2]{#1$\;${\small\rmfamily\@Roman{#2}}\relax}%

\newcommand\aap{{A\&A}}%
\newcommand\araa{{ARA\&A}}%
\newcommand\mnras{{MNRAS}}%
\newcommand\apj{{ApJ}}%
\newcommand\apjs{{ApJS}}%
\newcommand\apjl{{ApJ}}%
\newcommand\aj{{AJ}}%
\newcommand\pasp{{PASP}}%
\newcommand\memsai{{Mem.~Soc.~Astron.~Italiana}}%
\newcommand\procspie{{Proc.~SPIE}}%
\newcommand\nat{{\it Nature}}%

\setlist[enumerate]{noitemsep}
\setlist[enumerate,1]{leftmargin=*}
\setlist[itemize]{noitemsep}
\setlist[itemize,1]{leftmargin=*}
\setlist[description]{noitemsep}
\setlist[description,1]{leftmargin=*}
\setenumerate[0]{label=(\arabic*)}

\makeatother

\title[Integrated Abundances of PAndAS Clusters]{Integrated Light
Chemical Tagging Analyses of Seven M31 Outer Halo Globular Clusters
from the Pan-Andromeda Archaeological Survey\thanks{Based on
observations obtained with the Hobby-Eberly Telescope, which is a
joint project of the University of Texas at Austin, the Pennsylvania
State University, Stanford University,
Ludwig-Maximilians-Universit\"{a}t M\"{u}nchen, and
Georg-August-Universit\"{a}t G\"{o}ttingen.}}

\author[C. M. Sakari et al.]{Charli
M. Sakari$^{1}$\thanks{E-mail:sakaricm@u.washington.edu}
\thanks{Current address: Department of Astronomy, University of
Washington, Seattle WA 98195-1580, USA}
\thanks{Vanier Canada Graduate Scholar}, Kim A. Venn$^{1}$, Dougal
Mackey$ ^{2}$, Matthew D. Shetrone$^{3}$, 
\newauthor Aaron Dotter$^{2}$, Annette M.N. Ferguson$^{4}$, and Avon
Huxor$^{5}$\\
$^{1}$Department of Physics and Astronomy, University of Victoria,
Victoria, BC V8W 3P2, Canada\\
$^{2}$ Research School of Astronomy and Astrophysics, The Australian
National University, Weston, ACT 2611, Australia\\
$^{3}$McDonald Observatory, University of Texas at Austin, HC75 Box
1337-MCD, Fort Davis, TX 79734, USA\\
$^{4}$ Institute for Astronomy, University of Edinburgh, Royal
Observatory, Blackford Hill, Edinburgh EH9 3HJ, UK\\
$^{5}$ Astronomisches Rechen-Institut, Universit\"{a}t Heidelberg,
M\"{o}nchhofstra{\ss}e 12-14, D-69120 Heidelberg, Germany\\
}

\begin{document}

\maketitle

\label{firstpage}

\begin{abstract}
Detailed chemical abundances are presented for seven M31 outer halo 
globular clusters (with projected distances from M31 greater than 30
kpc), as derived from high resolution integrated light spectra taken
with the Hobby Eberly Telescope.  Five of these clusters were recently
discovered in the Pan-Andromeda Archaeological Survey (PAndAS)---this
paper presents the first determinations of integrated Fe, Na, Mg, Ca,
Ti, Ni, Ba, and Eu abundances for these clusters.  Four of the target
clusters (PA06, PA53, PA54, and PA56) are metal-poor
($[Fe/H]~<~-1.5$), $\alpha$-enhanced (though they are possibly less
$\alpha$-enhanced than Milky Way stars at the $1\sigma$ level), and
show signs of star-to-star Na and Mg variations.  The other three
globular clusters (H10, H23, and PA17) are more metal rich, with
metallicities ranging from $[\rm{Fe/H}] = -1.4$ to $-0.9$.
While H23 is chemically similar to Milky Way field stars, Milky Way
globular clusters, and other M31 clusters, H10 and PA17 have
moderately low [Ca/Fe], compared to Milky Way field stars and
clusters.  Additionally, PA17's high [Mg/Ca] and [Ba/Eu] ratios are
distinct from Milky Way stars, and are in better agreement with the
stars and clusters in the Large Magellanic Cloud (LMC).  None of the
clusters studied here can be conclusively linked to any of the
identified streams from PAndAS; however, based on their locations,
kinematics, metallicities, and detailed abundances, the most
metal-rich PAndAS clusters H23 and PA17 may be associated with the
progenitor of the Giant Stellar Stream, H10 may be associated with the
SW Cloud, and PA53 and PA56 may be associated with the Eastern Cloud.
\end{abstract}

\begin{keywords}
galaxies: individual(M31) --- galaxies: abundances --- galaxies: star
clusters: general --- globular clusters: general --- galaxies: evolution
\end{keywords}

\section{Introduction}\label{sec:Intro}
The Andromeda Galaxy (M31) is the nearest large neighbour to the Milky
Way (MW).  Its relative proximity offers a unique opportunity to study
the stellar populations of a nearby, large galaxy without being
hindered by interstellar extinction.  Though the stars in M31 are
considerably fainter than MW field stars, they can still be resolved
for photometric and low resolution spectroscopic observations while
entire globular clusters (GCs) can be targeted for integrated light
(IL) spectroscopy. Such observations contribute to studies of the
formation and evolution of M31 and its satellite system.

Deep surveys of M31's outer regions, such as the Isaac Newton
Telescope (INT) survey \citep{Ibata2001,Ferguson2002} and the
Pan-Andromeda Archaeological Survey (PAndAS;
\citealt{McConnachie2009}) have demonstrated that there is a
significant amount of coherent substructure in the outer halo; these
coherent streams indicate that M31's halo is currently being assembled
by accretion of dwarf satellites.  \citet{Ibata2014} present maps of
the outer halo, demonstrating that there are a multitude of streams,
plumes, and clouds of metal-poor stars, along with a significant
metal-rich population that is mostly located in the Giant Stellar
Stream (GSS) south of M31.  As much as 42\% of the most metal-poor
stars ($[\rm{Fe/H}] < -1.7$) lie in coherent substructure---for
metal-rich stars ($[\rm{Fe/H}] > -0.6$) this percentage rises to 86\%.
The rest of the outer halo stars are distributed in a ``smooth''
component.

The presence of metal-rich and metal-poor coherent streams implies
that M31 is currently accreting multiple dwarf satellites (see, e.g.,
\citealt{Johnston2008} for accretion simulations that lead to coherent
streams).  The locations and metallicities of the outer halo field
stars can be used to infer the nature of these dwarf galaxies.
\begin{enumerate}
\item The metal-rich GSS and the surrounding metal-rich features
indicate that M31 has accreted at least one fairly massive
($\sim$~LMC-mass) galaxy
\citep{Ibata2001,Ferguson2002,Ibata2007,Fardal2013,Ibata2014}.  Star
formation histories indicate that most of the metal rich stars formed
more than 5 Gyr ago, hinting that the progenitor may have been an
early-type dwarf \citep{Bernard2014}.
\item Intermediate [Fe/H] stars are also found in coherent
substructures, notably the SW Cloud \citep{Ibata2014,Bate2014}, which
has its own GCs (PA7, PA8, and PA14; \citealt{Mackey2013,Mackey2014})
and possibly its own \ion{H}{1} gas \citep{Lewis2013}.  Estimates
based on metallicity and total brightness suggest that the SW Cloud
progenitor was also a fairly massive dwarf galaxy (with a mass
comparable to the Fornax dwarf spheroidal; \citealt{Bate2014}).
\item The most metal-poor streams are gas-free (based on 21 cm
observations; \citealt{Lewis2013}) and likely originated in lower mass
dwarf spheroidal systems.
\end{enumerate}

The outer halo GCs also suggest that M31 has had a fairly active
accretion history.  Of the $\sim 60$ GCs discovered in PAndAS, many
appear to lie along stellar streams.  \citet{Mackey2010} demonstrated
that the positions of these GCs are correlated with the positions of
the streams, with a $<1$\% chance that these GCs are located along the
streams by chance. PA7 and PA8 are kinematically associated with each
other \citep{Mackey2013} and with the SW Cloud itself
\citep{Bate2014,Mackey2014}. \citet{Veljanoski2013,Veljanoski2014}
also show that many of the GCs which appear to lie along streams
could be associated, based on their kinematics.  \citet{Colucci2014}
have also demonstrated that at least one outer halo GC (G002) has
likely been accreted from a dwarf satellite.

Taken as a whole, the entire outer halo GC system is also different
from the MW's GC population.  For instance, \citet{Huxor2014} show
that M31 has more luminous and faint GCs than the MW---they suggest
that M31 could have acquired these extra GCs through accretion.
\citet{Veljanoski2014} also demonstrate that the entire GC system has
clear signs of rotation in both the GCs spatially associated with
streams and those that are unassociated with streams. This rotation
suggests either 1) that M31 experienced a merger with a galaxy large
enough to bring in a substantial fraction of the outer halo GCs, or 2)
that the parent dwarf galaxies were accreted from a preferred
direction such that their angular momenta were correlated, as is
presently seen for a separate plane of dwarf satellites
\citep{Ibata2013}.

Ultimately, the observations of outer halo stars and GCs indicate that
M31 has recently experienced a merger with at least one massive dwarf
and multiple lower mass dwarfs.  Detailed chemical abundances (from
high resolution spectroscopy) are an effective way to isolate
and identify coeval groups, a process known as ``chemical tagging''
(e.g., \citealt{Freeman,Mitschang2014}).   With abundances
of $\alpha$, iron-peak, and neutron capture elements, metal rich
($[\rm{Fe/H}] \ga -1.5$) stars and GCs from dwarf galaxies can be
distinguished from those in massive galaxies
(e.g. \citealt{Venn2004,Tolstoy2009,Ting2012}). Unlike GCs associated
with the MW and its dwarf satellites, however, distant extragalactic
targets must be observed through their integrated light (IL).  While
metallicity and $\alpha$-abundances can be determined from lower
resolution IL spectra (e.g.
\citealt{Schiavon2002,LeeWorthey2005,Puzia2006,Puzia2008,Caldwell2011}),
the detailed chemical abundances necessary for chemical tagging
analyses require high resolution IL spectroscopy (see, e.g.,
\citealt{McWB}; \citealt{Sakari2013}).  As in the MW
(\citealt{Cohen2004,Sbordone2005,Sakari2011}), if the M31 outer halo
GCs originated in lower mass dwarf galaxies then they could have
distinct abundance patterns from MW field stars, depending on
metallicity.

High resolution IL spectroscopy has been tested extensively on
Galactic GCs (\citealt{McWB}; \citealt{Cameron2009};
\citealt{Sakari2013,Sakari2014}), demonstrating that
\begin{enumerate}
\item IL abundances trace the individual stellar abundances and
represent cluster averages when abundances do not vary between stars
\item Abundance determinations from high resolution IL spectroscopy
are more precise than from lower resolution studies
\item Certain abundance ratios are very stable to uncertainties in the
underlying stellar populations (e.g. interloping field stars,
uncertainties in cluster age, microturbulence relations, etc.), such
as [Ca/Fe]
\item Partially resolved photometry down to the horizontal branch (HB)
reduces systematic uncertainties by constraining [Fe/H] and HB
morphology.
\end{enumerate}
Thus, a high resolution IL spectroscopic analysis of M31 GCs can
provide precise and accurate abundances that will be suitable for
chemical tagging.   High resolution IL spectroscopic techniques have
been applied to nearby GCs in the Local Group (in M31,
\citealt{Colucci2009,Colucci2014}; the LMC, \citealt{Colucci2011a} and
\citealt{Colucci2012}; and in dwarf galaxies, \citealt{Colucci2011b})
and in the large elliptical galaxy NGC~5128 \citep{Colucci2013}.  At
the distance of M31, the size and brightness of its GCs make them
ideal targets for high resolution IL spectroscopy.

This paper presents a high resolution IL chemical tagging analysis of
seven outer halo M31 GCs, five of which were first observed in PAndAS.
This is the first chemical analysis for these five PAndAS clusters, at
any resolution.  Section \ref{sec:Data} presents the data and the
analysis methods, while Section \ref{sec:ModelAtms} describes the
process for generating synthetic stellar populations for each
cluster. The IL abundances are given in Section \ref{sec:Abunds}, and
the implications for the formation of M31's outer halo are then
discussed in Section \ref{sec:Discussion}.  Finally, the results are
summarized in Section \ref{sec:Conclusions}.

\section{Observations and Analysis Methods}\label{sec:Data}

\subsection{Target Selection}\label{subsec:Targets}
\begin{figure*}
\begin{center}
\includegraphics[scale=1.0]{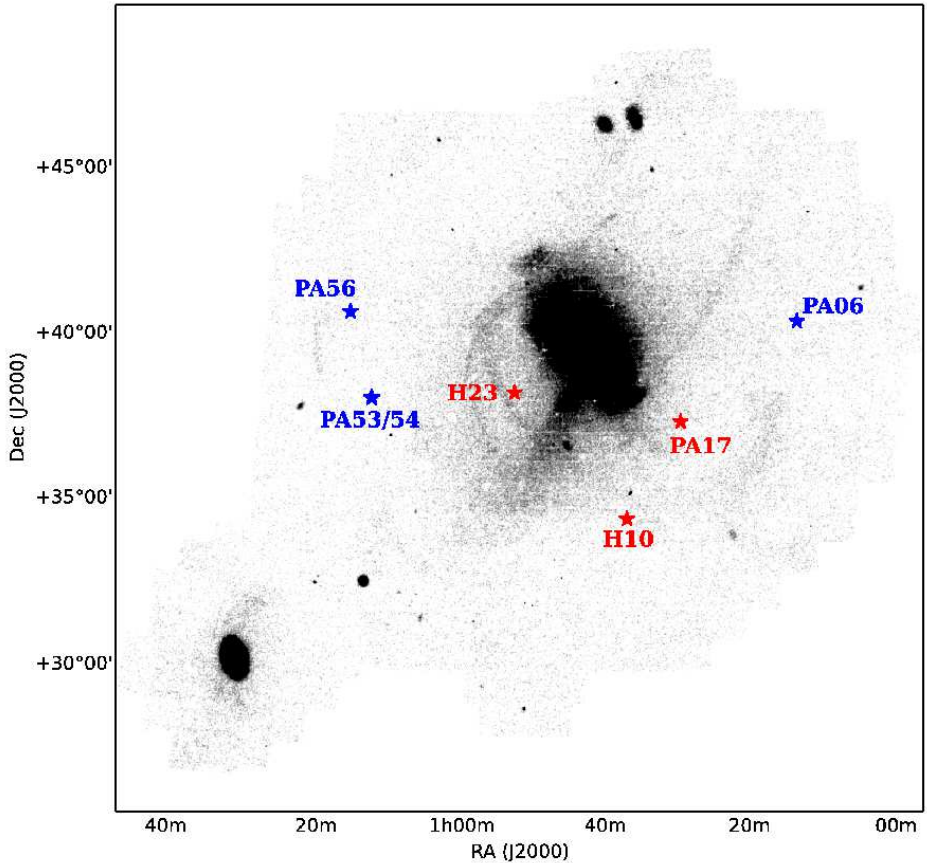}
\caption[PAndAS cluster locations.]{Locations of the target PAndAS
clusters on a metal-poor density map of the full PAndAS footprint.
The locations of stellar streams and satellite galaxies are obvious as
dark regions.  The red stars show the metal-rich ($[\rm{Fe/H}] \ga
-1.5$; see Section \ref{sec:Abunds}) PAndAS clusters observed in this
paper, while the blue stars show the metal-poor GCs.  Note that PA53
and PA54 are very close together in projection and share a single
point in the plot.\label{fig:PAndASClusterLocations}}
\end{center}
\end{figure*}

The primary goal of this high resolution spectroscopic investigation
of M31 GCs is to study the nature of the stars and clusters in M31's
outer halo. The targets were therefore restricted to GCs with
projected distances from the centre of M31 that are $R_{\rm{proj}} >
30$ kpc; priority was given to the GCs with the largest projected
radii.  Observational constraints required that the targets were
sufficiently bright to observe in a reasonable amount of time and to
avoid stochastic sampling issues---as a result, the target GCs are on
the bright end of the M31 outer halo GC distribution (see
\citealt{Huxor2014}).  Seven outer halo GCs were targeted; their
properties are summarized in Table \ref{table:PAndASTargets}. Two of
these target clusters (H10 and H23) were discovered by
\citet{Huxor2008}; the other five were discovered in the PAndAS
programme \citep{Huxor2014}.  The locations of these GCs on a density
map of metal-poor outer halo stars (from M. Irwin, {\it private
communication}) is shown in Figure
\ref{fig:PAndASClusterLocations}. Note that none of the targets fall
precisely onto any of the major stellar streams or overdensities in
the outer halo.  Nonetheless, PA56 sits on a faint north-western
extension of the Eastern Cloud, and is likely associated with this
feature (Mackey et al. 2015, {\it in prep.}).  While H23 has been
tentatively linked to Stream D based on its relative proximity in
projection \citep{Mackey2010}, its radial velocity renders a genuine
association unlikely \citep{Veljanoski2014}.

\begin{table*}
\centering
\begin{minipage}{165mm}
\begin{center}
\caption{Properties of the target PAndAS clusters.\label{table:PAndASTargets}}
  \newcolumntype{d}[1]{D{.}{.}{#1} }
  \begin{tabular}{@{}lccd{1}d{1}ccc@{}}
  \hline
Cluster & RA (J2000) & Dec (J2000) &
\multicolumn{1}{c}{$V_{\rm{int}}$} &
\multicolumn{1}{c}{$R_{\rm{proj}}^{a}$} & Photometric$^{b}$ & Spatial
Association & References\\
        &            &             &                &
\multicolumn{1}{c}{(kpc)}           & [Fe/H]       & with Stream? & \\
 \hline
H10   & $00^{\rm{h}}35^{\rm{m}}59.7^{\rm{s}}$ & $+35^{\circ}41\arcmin03\arcsec.6$ & 15.7 & 78.5 & -1.84 & N & 1, 2\\
H23   & $00^{\rm{h}}54^{\rm{m}}25.0^{\rm{s}}$ & $+39^{\circ}42\arcmin55\arcsec.5$ & 16.8 & 37.1 & -1.54 & Stream D? & 1, 2\\
PA06  & $00^{\rm{h}}06^{\rm{m}}12.0^{\rm{s}}$ & $+41^{\circ}41\arcmin21\arcsec.0$ & 16.5 & 93.7 &   MP  & N & 2, 3, 4\\
PA17  & $00^{\rm{h}}26^{\rm{m}}52.2^{\rm{s}}$ & $+38^{\circ}44\arcmin58\arcsec.1$ & 16.3 & 53.9 &    ?$^{c}$ & N & 2, 3\\
PA53  & $01^{\rm{h}}17^{\rm{m}}58.4^{\rm{s}}$ & $+39^{\circ}14\arcmin53\arcsec.2$ & 15.4 & 95.9 &   MP  & N & 2, 3, 4\\
PA54  & $01^{\rm{h}}18^{\rm{m}}00.1^{\rm{s}}$ & $+39^{\circ}16\arcmin59\arcsec.9$ & 15.9 & 95.8 &   MP  & N & 2, 3, 4\\
PA56  & $01^{\rm{h}}23^{\rm{m}}03.5^{\rm{s}}$ & $+41^{\circ}55\arcmin11\arcsec.0$ & 16.8 &103.3 &   MP  & N & 2, 3, 4\\
 & & & & & & \\
\hline
 & & & & & & & \\
\end{tabular}\\
\end{center}
\end{minipage}
\raggedright {\bf References: } 1 = \citet{Mackey2007}, 2 =
\citet{Veljanoski2014}, 3 = \citet{Huxor2014}, 4 = Mackey et al.,
\textit{in prep.}.\\
$^{a}$ Projected distances are from the centre of M31.\\
$^{b}$ Photometric metallicity estimates are used to constrain the
parameters of the input isochrones; see Section \ref{sec:ModelAtms}.
[Fe/H] estimates are based on Galactic GC fiducial fits.  MP indicates
a metal-poor cluster ($[\rm{Fe/H}] \la -1.5$); fiducial fits will be
given in Mackey et al., \textit{in prep.}.\\
$^{c}$ PA17 does not have an {\it HST} CMD, and there is therefore no
{\it a priori} information about its metallicity or HB morphology.\\
\end{table*}

\citet{Sakari2014} present detailed tests of systematic abundance
errors that occur because of uncertainties in the underlying stellar
populations.  These tests include:
\begin{itemize}
\item Uncertainties in isochrone parameters such as age and [Fe/H]
\item Properties of evolved horizontal branch (HB) and asymptotic
giant branch (AGB) stars, particularly the HB morphology
\item Assumptions about the stellar populations (e.g. microturbulence
relations, initial mass functions, mass segregation)
\item Unusual stars like interloping field stars, long period
variables, etc.
\end{itemize}
These errors can be mitigated if GC properties are constrained with
photometry or lower resolution analyses over a broader wavelength
range.  Observing priority was therefore given to those targets with
partially resolved photometry.  Six of the seven target GCs have
\textit{Hubble Space Telescope} ({\it HST}) photometry down to the HB
(\citealt{Mackey2007},Mackey et al. 2015, \textit{in prep.}).  This
photometry is used to place constraints on the best fitting isochrones
(see Section \ref{sec:ModelAtms}).

\subsection{Observations and Data Reduction}\label{subsec:Observations}
The targets were observed with the Hobby-Eberly Telescope
(HET; \citealt{HETref,HETQueueref}) at McDonald Observatory in Fort
Davis, TX in 2011 and early 2012.  The High Resolution Spectrograph
(HRS; \citealt{HRSref}) was utilized with the 3$\arcsec$ fibre and a
slit width of 1$\arcsec$, yielding an instrumental spectral resolution
of $R~=~30,000$. With the 600 gr/mm cross disperser set to a central
wavelength of 6302.9 \AA, wavelength coverages of $\sim 5320-6290$~\AA
\hspace{0.025in} and $\sim 6360-7340$~\AA \hspace{0.025in} were
achieved in the blue and the red, respectively.  The 3$\arcsec$ fibre
provided coverage of the clusters past their half-light radii; the
additional sky fibres (located 10$\arcsec$ from the central object
fibre) provided simultaneous observations for sky subtraction.
Exposure times were calculated to obtain a total S/N~=~80 (per
resolution element), although not all targets received sufficient
time to meet this goal. The details of the observations are shown in
Table \ref{table:PAndASObservations}.

\begin{table*} 
\centering
\begin{minipage}{165mm}
\begin{center}
\caption{PAndAS cluster observations.\label{table:PAndASObservations}}
  \newcolumntype{d}[1]{D{,}{\pm}{#1} }
  \begin{tabular}{@{}llccccd{3}d{3}d{3}@{}}
  \hline
Cluster & Observation & Exposure & S/N$^{a}$  & S/N$^{a}$  & $r_{\rm{ILS}}^{b}$ & \multicolumn{1}{c}{$v_{\rm{helio, obs}}$} & \multicolumn{1}{c}{$v_{\rm{helio, lit}}$} & \multicolumn{1}{c}{$\sigma_{\rm{obs}}$} \\
        & Dates       & Time (s) & (5500 \AA) & (7000 \AA) & ($r_h$)            & \multicolumn{1}{c}{(km/s)} & \multicolumn{1}{c}{(km/s)} & \multicolumn{1}{c}{(km/s)} \\
  \hline
H10  & 2011 Jan 2, 10, 11, 22, 23, 28, 30              & 19180 & 82 & 140 & 2.3 & -351.9,1.5 & -352,9  & 6.6,0.4 \\
 & & & & &  & & & \\
H23  & 2011 Jul 5, 7, 10, 11, Aug 2, 4, Sep 23         & 16050 & 65 & 82 & 4.1 & -373.3,0.1 & -377,11 & 6.2,0.4 \\
 & & & & & & &  & \\
PA06 & 2011 Sep 29, Oct 1, 2, 4, 6, 17, 18, 19, 21, 23 & 29278 & 65 & 105 & 2.6 & -341.4,0.7 & -327,15 & 5.6,0.4 \\
 & & & & & & &  & \\
PA17 & 2012 Jan 16, 18, 21                             & 8100  & 41 & 50  & 2.6 & -260.0,1.0 & -279,15 & 6.1,0.5 \\
 & & & & & & &  & \\
PA53 & 2011 Aug 1, Sep 20, 26                          & 8100  & 115 & 148 & 2.7 & -270.8,0.9 & -253,10 & 12.0,0.4\\
 & & & & & & &  & 	\\			               
PA54 & 2011 Aug 25, Sep 24, 25, 27                     & 10800 & 82 & 130 & 2.2 & -344.9,0.8 & -336,8  & 7.5,0.4 \\
 & & & & & & &  & \\
PA56 & 2011 Oct 7, 17, 18, 20, 21, 23, 24, 30,         & 30623 & 65 & 82 & 2.4 & -241.4,1.7 & -239,8  & 6.4,0.4 \\
     & \phantom{2011} Nov 19, Dec 30,  &       &    &    &     & & &\\
     & 2012 Feb 10, 11  &       &    &    &     & & &\\
 & & & & & & &  & \\
\hline
 & & & & & & &  & \\
\end{tabular}
\end{center}
\end{minipage}\\
\medskip
\raggedright {\bf References: } Literature radial velocities are from
\citet{Veljanoski2014}\\
\raggedright $^{a}$S/N ratios are per resolution element, and assume
that there are 2.7 pixels per resolution element for HRS.\\
$^{b}$The coverage radii are based on the half-light radii in
\citet{Tanvir2012} and \citet{Huxor2014}---all GCs are covered past
their half-light radii.\\
\end{table*}

Data reduction was performed in in the Image Reduction and Analysis
Facility program (IRAF).\footnote{IRAF is distributed by the National
Optical Astronomy Observatory, which is operated by the Association of
Universities for Research in Astronomy, Inc., under cooperative
agreement with the National Science Foundation.}  As in
\citet{Sakari2013}, bias removal was not performed to avoid adding
noise to the individual spectra, and variance weighting was used during
aperture extraction to remove cosmic rays.  Sky spectra (from the
separate sky fibres) were replaced with continuum fits with the
emission lines added back in (according to the sky line
identifications from the UVES Quality Control sky spectrum
website\footnote{\url{http://www.eso.org/observing/dfo/quality/UVES/pipeline/sky_spectrum.html}})
and were then subtracted from the object spectra.  Telluric standards
were also observed for removal of atmospheric absorption features.

In order to avoid removing broad, blended features, the target
spectral orders were normalized with continuum fits to an
extremely metal-poor star (CS29502-092; see
\citealt{Sakari2013}). Low-order polynomial fits were then necessary
to fully normalize the target spectra.  The individual observations
were cross-correlated with a high resolution, high S/N Arcturus
spectrum (from the Arcturus
Atlas\footnote{\url{ftp://ftp.noao.edu/catalogs/arcturusatlas/}};
\citealt{Hinkle2003}) to determine radial velocities.  Heliocentric
velocities were determined for each individual observation; the final,
averaged heliocentric velocities are shown in Table
\ref{table:PAndASObservations}.  The quoted errors represent the
dispersion between observations.

\begin{figure*}
\begin{center}
\includegraphics[trim=0.in 0.75in 0.2in 0in,scale=0.75]{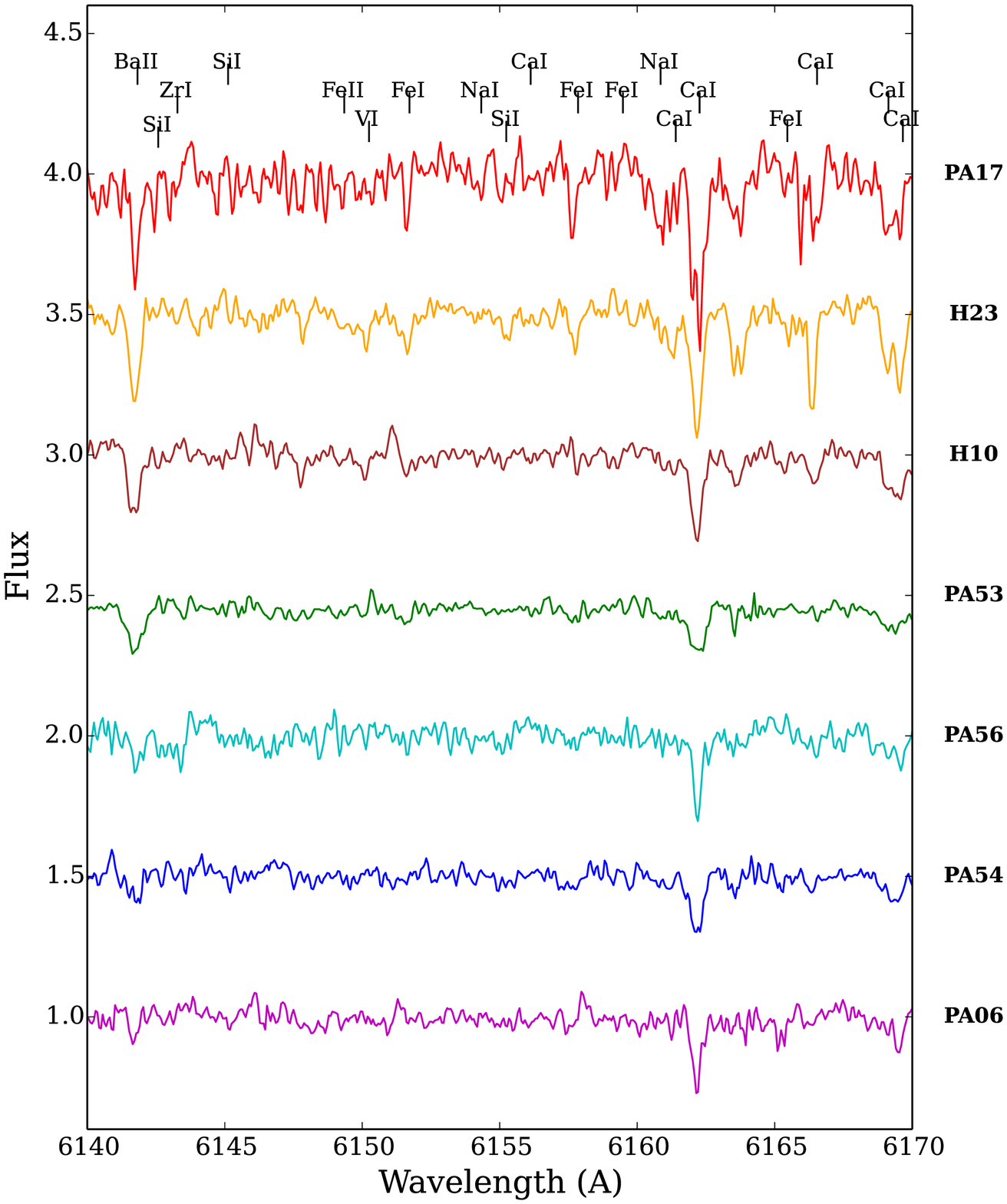}
\caption{IL spectra of the PAndAS GCs, arranged by
metallicity. Spectral lines used in typical RGB stellar analyses are
identified.\label{fig:PAndASGCSpectra}}
\end{center}
\end{figure*}

The individual, rest-frame spectra were combined with average
sigma-clipping rejection routines to remove any remaining cosmic
rays; each spectrum was weighted by flux during the combination.  The
combined spectra were once again cross-correlated with the Arcturus
template spectrum to determine cluster dispersions (see
\citealt{Sakari2013} for a description of how this is done).  These
velocity dispersions are also shown in Table
\ref{table:PAndASObservations}. Examples of the final, combined
spectra are shown in Figure \ref{fig:PAndASGCSpectra}.

\subsection{Line List and EW Measurements}\label{subsec:DAOSPEC}
For these chemical tagging analyses, spectral lines of Fe, Na, Mg, Ca,
Ti, Ni, Ba, and Eu are utilized.  \citet{Sakari2013,Sakari2014}
demonstrated that for Galactic GCs reliable Fe, Ca, Ti, Ni, and Ba
abundances can be determined from equivalent widths (EWs), while Na,
Mg, and Eu abundances can be determined with spectrum syntheses.  Note
that in this analysis the Ba lines are synthesized instead of using
EWs, due to the lower S/N ratios of the target spectra.

EWs for Fe, Ca, Ti, and Ni lines were measured with the program
DAOSPEC\footnote{DAOSPEC has been written by P.B. Stetson for the
Dominion Astrophysical Observatory of the Herzberg Institute of
Astrophysics, National Research Council, Canada.} \citep{DAOSPECref}.
\citet{Sakari2013} verified that DAOSPEC measurements of IL spectral
lines compare well with measurements from other methods for GCs with a
range of velocity dispersions.  Given the low S/N of some of the
PAndAS targets, many of the DAOSPEC EWs were verified or refined by
hand.  IL spectral lines have contributions from stars with a variety
of line strengths---in particular, IL lines with EWs $\ga 100$ m\AA
\hspace{0.025in} can have contributions from bright red giant stars
with EWs $\sim 200$ m\AA.  These strong lines are difficult to model
(see the discussions in Paper I and \citealt{McWilliam1995b}) and are
typically removed from individual stellar analyses.  For this reason,
as a conservative estimate, all lines stronger than $\sim110$ m\AA
\hspace{0.025in} were removed from this analysis unless absolutely
necessary.

Table \ref{table:LineList} shows the spectral lines that were used to
derive abundances in the analysis, their atomic data, and the measured
IL EWs for each cluster. The line list is based on the standard IL
lines lists of \citet{McWB} and \citet{Colucci2009}, with supplements
from the Red Giant Branch (RGB) stellar line lists of
\citet{Sakari2011} and \citet{Venn2012}. Lines that were measured with
spectrum syntheses are indicated.  The full line lists to synthesize
all lines in a 10 \AA \hspace{0.025in} window (for spectrum syntheses)
are the ones used in \citet{Sakari2013}: they consist of the EW line
list, with supplements from the Vienna Atomic Line
Database\footnote{\url{http://www.astro.uu.se/~vald/php/vald.php}}
(VALD; \citealt{Kupka2000}), the Kurucz
database,\footnote{\url{http://kurucz.harvard.edu/linelists.html}} and
the National Institute of Standards and Technology
(NIST)\footnote{\url{http://www.nist.gov/index.html}} database.
Molecular features from the Kurucz database were included when
indicated in the Arcturus Atlas (for CH, CN, and MgH).  
Isotopic and hyperfine structure (HFS) components for the
\ion{Ba}{2} lines are from \citet{McW1998} while the \ion{Eu}{2}
components are from \citet{LawlerA} and \citet{LawlerB}.  The HFS
components are included in the syntheses; while they do not
significantly affect the strengths of the lines, they can affect the
line profiles.

\begin{table*}
\centering
\begin{minipage}{165mm}
\begin{center}
\caption{The Line List.$^{a}$\label{table:LineList}}
  \begin{tabular}{@{}lcccccccccc@{}}
  \hline
$\lambda$ & Element & E.P. & log gf & \multicolumn{7}{l}{Equivalent width (m\AA)}\\
(\AA) & & (eV) & & H10 & H23 & PA06 & PA17 & PA53 & PA54 & PA56\\
\hline
5324.191 & \ion{Fe}{1} & 3.21  & -0.103 & -    & -    & 80.0 & -    & 83.1 & 79.4 & 80.0 \\
5339.937 & \ion{Fe}{1} & 3.27  & -0.720 & 81.0 & -    & -    & -    & -    & -    & 54.8 \\
5367.476 & \ion{Fe}{1} & 4.42  &  0.443 & 77.9 & 93.9 & 46.0 & -    & 68.0 & 50.0 & 53.2 \\
5369.974 & \ion{Fe}{1} & 4.37  &  0.536 & -    & -    & 50.0 & -    & -    & -    & 59.4 \\
5371.501 & \ion{Fe}{1} & 0.96  & -1.644 & -    & -    &126.0 & -    & -    & -    & -\\
\hline
\end{tabular}
\end{center}
\end{minipage}\\
\medskip
\raggedright {\bf Notes: } Equivalent widths were measured in DAOSPEC;
all strong lines were checked and refined in \textit{splot}.  IL lines
stronger than 110 m\AA \hspace{0.025in} were not included in the
analysis.\\
$^{a}$Table \ref{table:LineList} is published in its
entirety in the electronic edition of \textit{Monthly Notices of the
Royal Astronomical Society}. A portion is shown here for guidance
regarding its form and content.\\
\end{table*}

\subsection{Solar Abundances}\label{subsec:Sun}
\citet{Sakari2013,Sakari2014} presented measurements of the solar
spectrum from the \citet{Kurucz2005} solar flux
atlas.\footnote{\url{http://kurucz.harvard.edu/sun.html}}  The [Fe/H]
and [X/Fe] ratios in this paper were calculated {\it line by line}
using the solar abundances derived from those EWs and spectrum
syntheses.  When the solar lines were stronger than 150~m\AA,
\citet{Asplund2009} solar abundances were adopted for those lines.

\section{Underlying Stellar Populations}\label{sec:ModelAtms}
The first step in an IL analysis is to determine the atmospheric
parameters of the stars in the underlying populations.  The central
regions of the PAndAS GCs cannot be resolved with {\it HST}, and high
quality CMDs cannot be obtained for the regions included in the IL
spectra---isochrones must therefore be used to model the stellar
populations.  Synthetic populations are generated with the BaSTI
isochrones \citep{BaSTIref,Cordier2007} with extended AGBs and mass
loss parameters of $\eta = 0.2$, assuming a \citet{Kroupa2002} IMF and
the total cluster magnitudes from \citet{Huxor2014}.  The BaSTI
isochrones are utilized because they extend all the way through the
AGB phase---\citet{Sakari2014} show that differences between isochrone
sets are insignificant for the abundances derived in this
paper. The synthetic Hertzsprung Russell Diagram (HRD) populations are
binned into boxes, each with 3.5\% of the total flux. The boxes are
then assigned corresponding Kurucz model
atmospheres\footnote{\url{http://kurucz.harvard.edu/grids.html}}
\citep{KuruczModelAtmRef}, where the grid values are interpolated to a
specific box's $T_{\rm{eff}}$ and $\log g$.  When the [Ca/Fe] ratio
is supersolar, $\alpha$-enhanced model atmospheres are used.
Microturbulent velocities are determined via an empirical relation
between $\xi$ (in km s$^{-1}$) and $\log g$ that fits the Sun and
Arcturus (see \citealt{McWB}).

The abundances of the \ion{Fe}{1} lines (from the EWs in Table
\ref{table:LineList}) are used to constrain the best-fitting
isochrone's single age and metallicity.  As in \citet{McWB} and
\citet{Colucci2009,Colucci2011a,Colucci2013}, the adopted isochrone is
the one that leads to the flattest trends in \ion{Fe}{1} abundance
with wavelength, reduced EW (REW),\footnote{$\rm{REW} =
\rm{EW}/\lambda$} and excitation potential (EP); these criteria are
adopted from analyses of individual stars (see \citet{McWB}).

\subsection{Clusters with {\it HST} Photometry}\label{subsec:HST}
\begin{figure*}
\begin{center}
\centering
\subfigure[H10]{\includegraphics[scale=0.35]{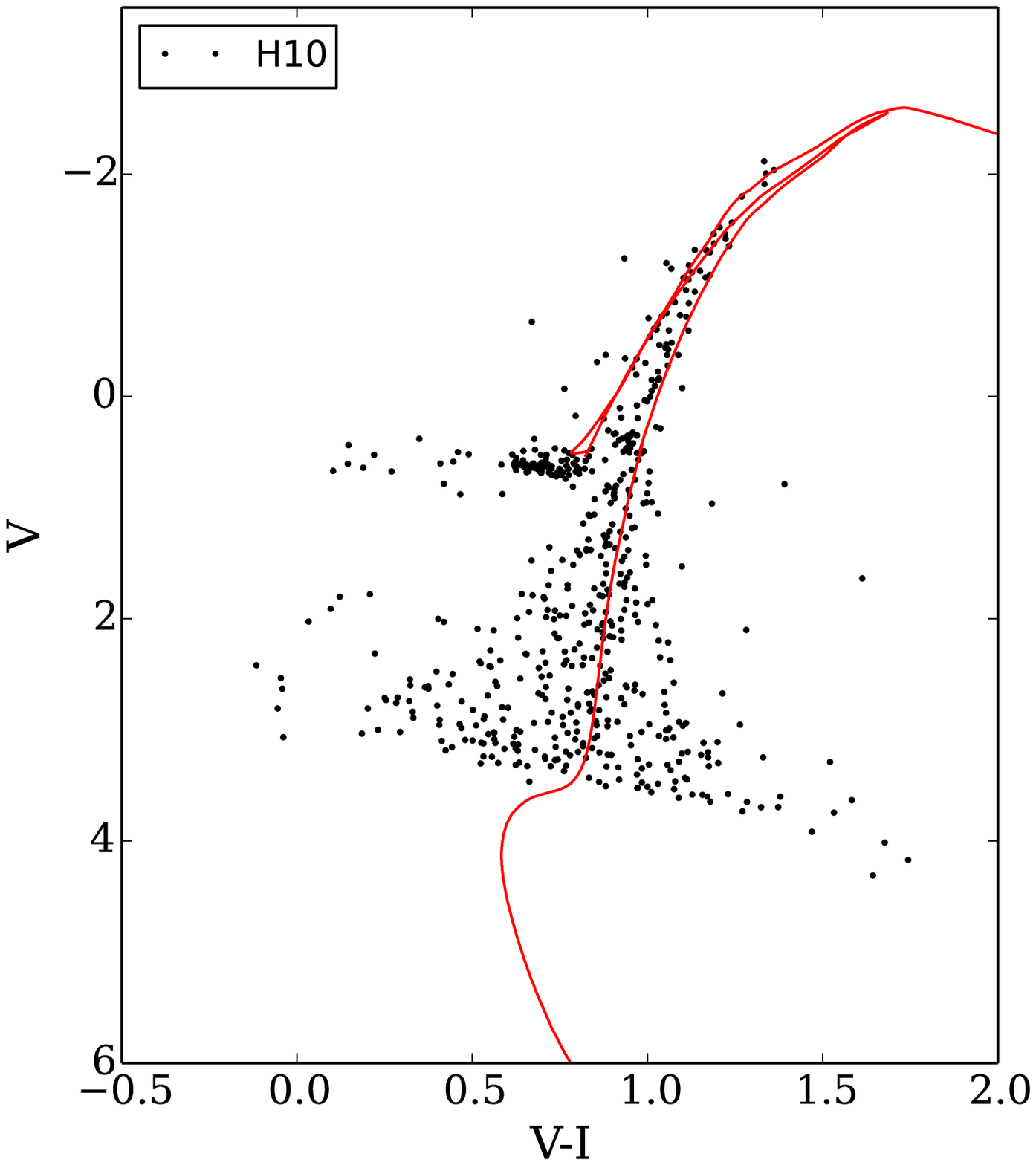}\label{fig:H10CMD}}
\subfigure[H23]{\includegraphics[scale=0.35]{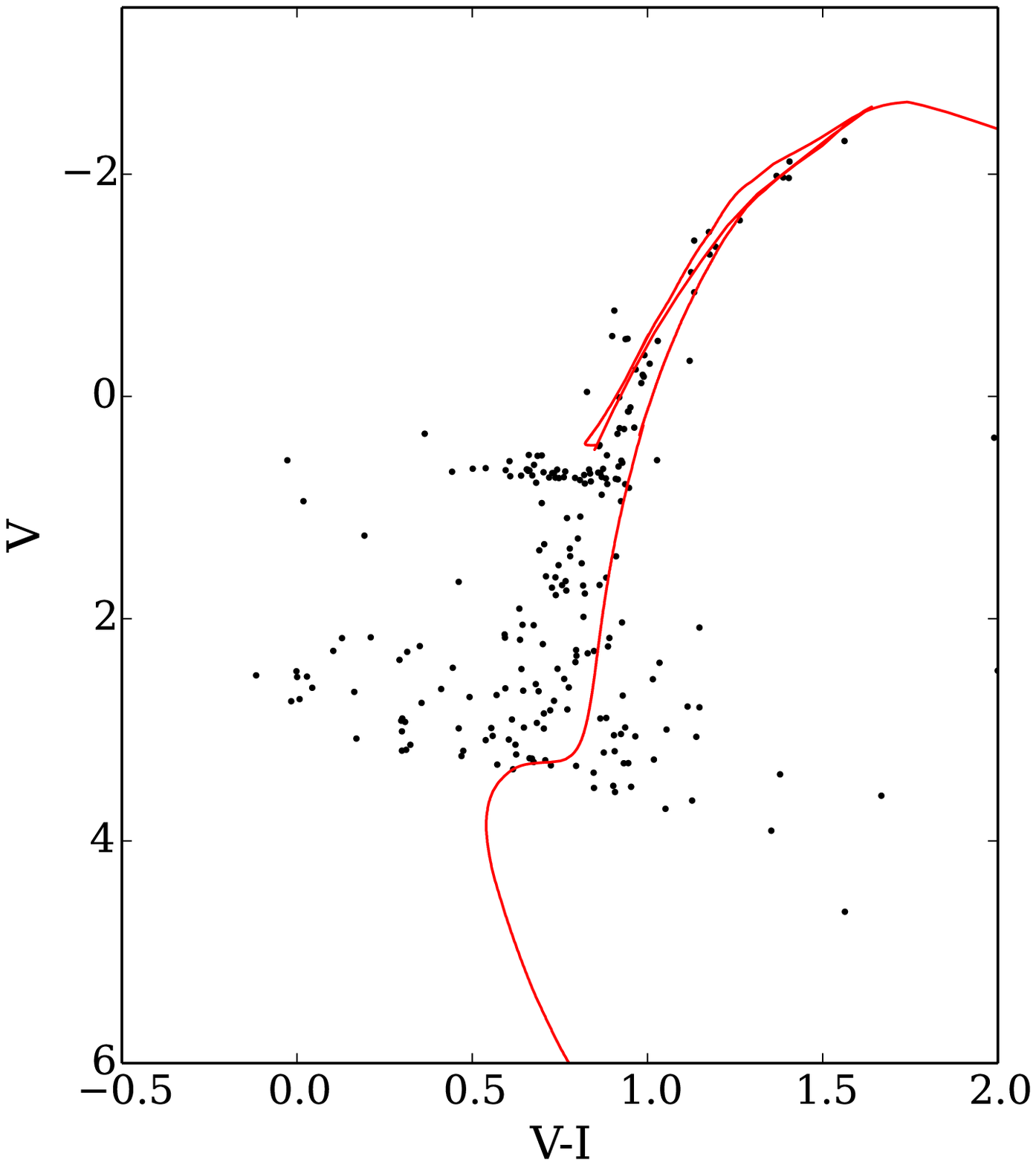}\label{fig:H23CMD}}
\subfigure[PA06]{\includegraphics[scale=0.35]{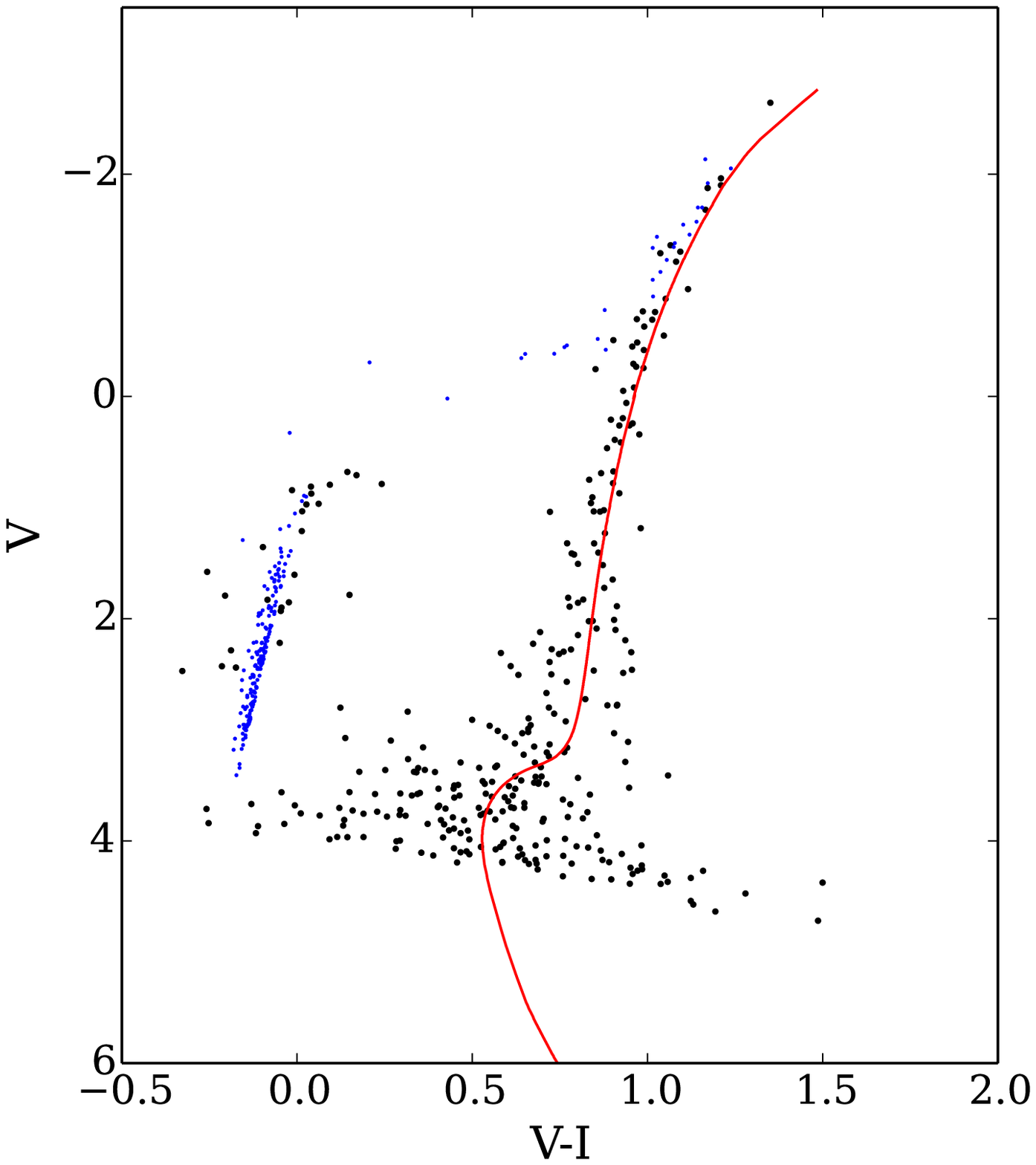}\label{fig:PA06CMD}}
\subfigure[PA53]{\includegraphics[scale=0.35]{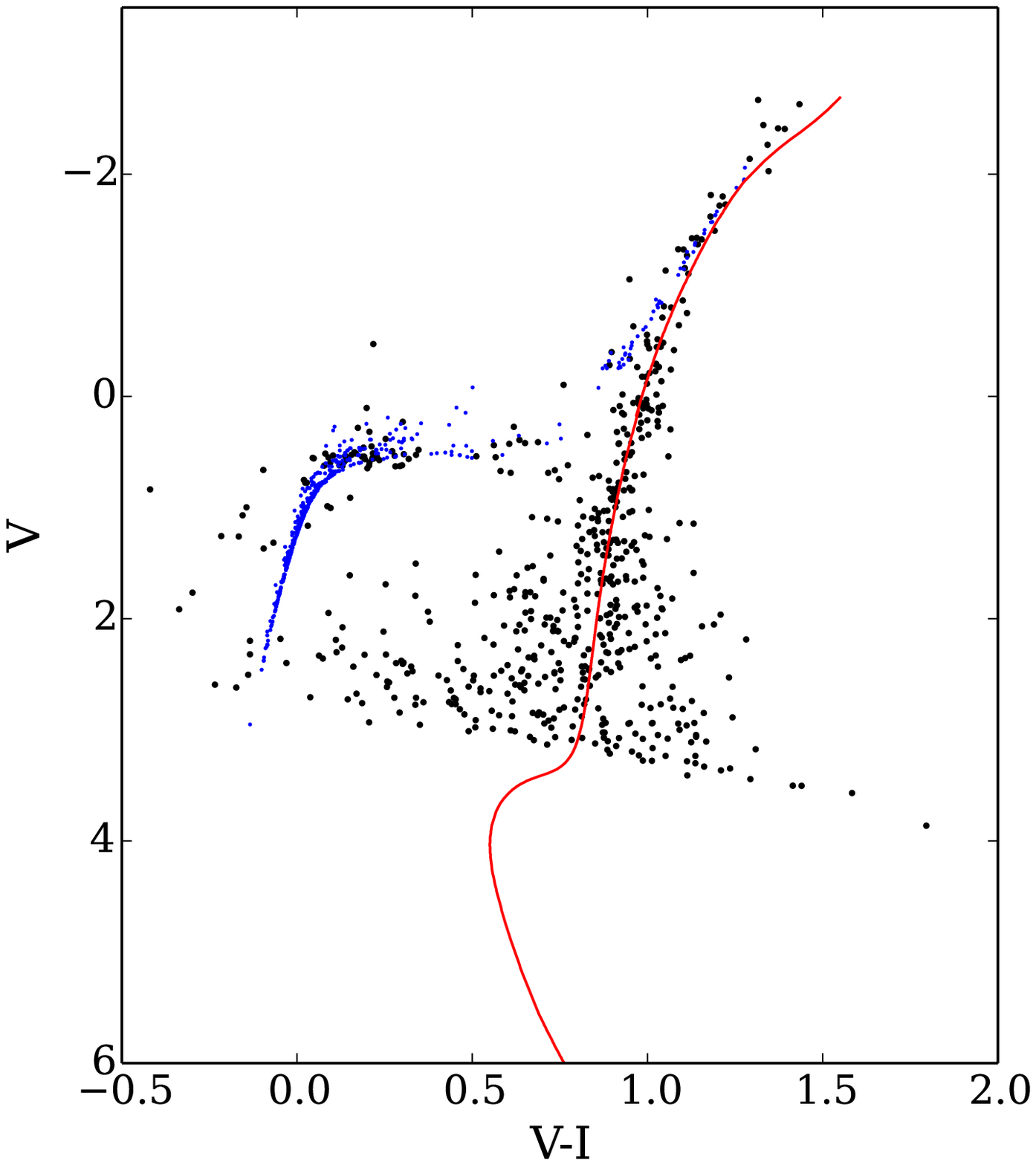}\label{fig:PA53CMD}}
\subfigure[PA54]{\includegraphics[scale=0.35]{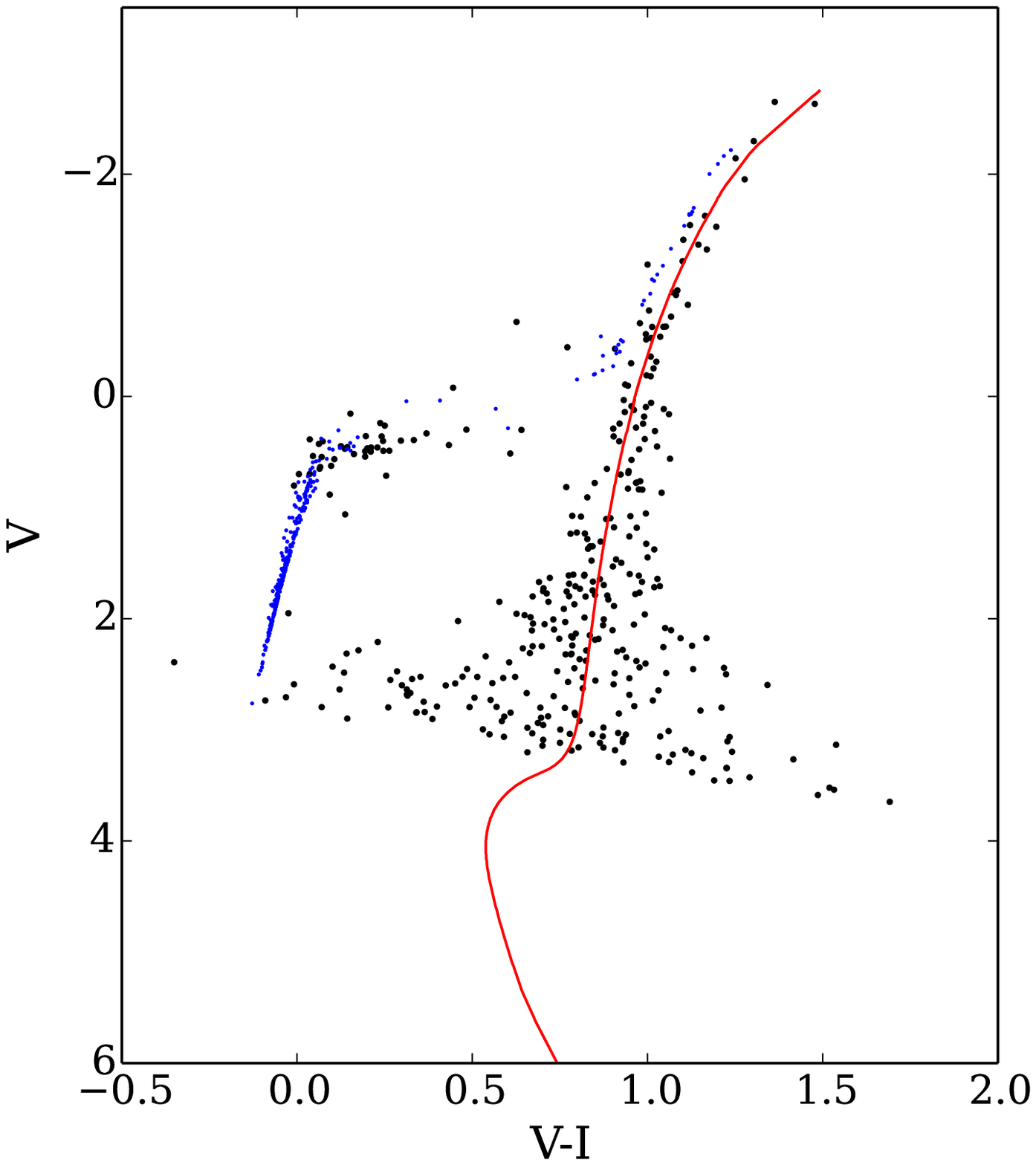}\label{fig:PA54CMD}}
\subfigure[PA56]{\includegraphics[scale=0.35]{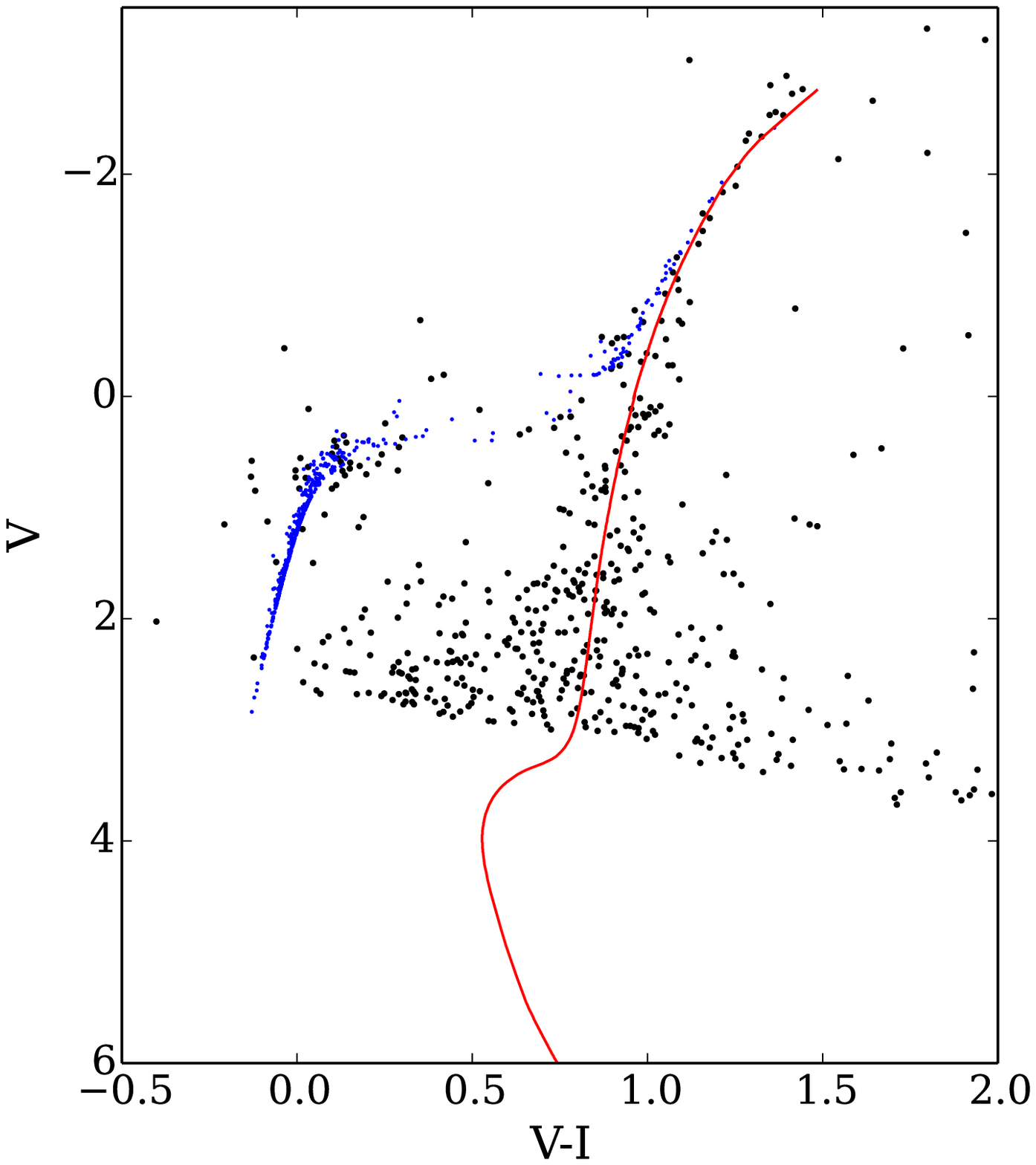}\label{fig:PA56CMD}}
\caption{{\it HST} CMDs of the six partially resolved PAndAS GCs.  The
spectroscopically determined best-fitting isochrones are shown (in
red) on the CMDs to illustrate how the synthetic HRDs are
populated.  For the blue HB clusters PA06, PA53, PA54, and PA56,
synthetic HBs (shown in blue) were used instead of the default
HBs.\label{fig:CMDs}}
\end{center}
\end{figure*}

The stars in the {\it outer} regions of six of the targets clusters
{\it have} been resolved with {\it HST}, enabling constraints to be
placed on the GC [Fe/H] and HB morphology.\footnote{Rough constraints
can also be placed on cluster age, specifically whether the cluster is
older than $\sim 4$ Gyr; additionally, the presence of RR Lyrae stars
would constrain the age to $>10$ Gyr.}  This photometry is not used to
identify the isochrone that best fits the resolved CMD, but is only
used to constrain the possible isochrones and HB morphology.  These
constraints help to drastically reduce the systematic errors in the IL
abundances (see \citealt{Sakari2014}).  For the blue HB clusters
(PA06, PA53, PA54, and PA56; see Figure \ref{fig:CMDs}), synthetic
BaSTI
HBs\footnote{\url{http://basti.oa-teramo.inaf.it/BASTI/WEB_TOOLS/HB_SYNT/index.html}}
are used instead of the default HBs.  \citet{Sakari2014} show that as
long as the HBs are modelled approximately correctly the integrated
abundances will not be systematically affected.

The parameters of the spectroscopically-determined isochrones are
listed in Table \ref{table:HRDs} and the isochrones themselves are
shown on top of the observed CMDs in Figure \ref{fig:CMDs} (for the
six clusters with  {\it HST} photometry).  Again, {\it these
isochrones were not derived based on fits to the observed CMDs}, but
were determined by minimizing trends in the \ion{Fe}{1} line
abundances. The fact that the isochrones agree well with the CMDs
confirms the validity of these high resolution IL techniques.

The slopes of the trends in \ion{Fe}{1} abundances are given in Table
\ref{table:PAndASSlopes}.  Again, solutions were selected by
simultaneously minimizing trends in \ion{Fe}{1} abundance with
wavelength, REW, and EP.  All REW and EP trends for the PAndAS
clusters are flat within their $1\sigma$ errors. The line-to-line
dispersion can still be quite large, which leads to some uncertainty
in isochrone age ($\sim~5$~Gyr), though this has a minimal effect on
the integrated abundances \citep{Sakari2014}.  Also note that the
BaSTI isochrones have large spacings in metallicity ($\sim 0.2-0.3$
dex, depending on [Fe/H]), preventing exact metallicities from being
adopted for the isochrones.  Comparisons with the integrated colours
are shown in Table \ref{table:HRDs}.  The differences for the six
partially resolved GCs are all $\la 0.05$ magnitudes (which is the
approximate uncertainty in the observed colours from
\citealt{Huxor2014}); this indicates that the populations are at least
reasonably well modelled. H10 and H23's slightly redder synthetic
colours may be due to the fact that the default HBs are slightly too
red (see Figures \ref{fig:H10CMD} and \ref{fig:H23CMD}), while the
inconsistencies in the other GCs might be due to incorrect ages or
slight discrepancies in the RGBs, HBs, or AGBs.

\subsection{PA17: The Cluster without {\it HST} Photometry}\label{subsec:NoHST}
Since PA17 does not yet have a high quality {\it HST} CMD there is
little {\it a priori} information available to constrain its age,
[Fe/H], or HB morphology.  As a result the isochrone default HBs were
used.  The flattest slopes occur for a relatively metal-rich isochrone,
with $[\rm{Fe/H}]\sim -1$, and an age of 12 Gyr.  Given PA17's [Fe/H]
and red integrated colour ($(V-I)_0 = 1.14$, which is redder than
H10 and H23; \citealt{Huxor2014}), the default BaSTI HB (which is red)
is likely to be sufficient, and there abundance offsets as a result of
the uncertain HB morphology should be negligible.  With the default
HBs, the \ion{Fe}{1} abundances imply a best-fitting age and [Fe/H] of
12~Gyr and -1.01, respectively, for PA17.

\begin{table}
\centering
\begin{center}
\caption{Parameters of the ``best-fitting'' HRDs and synthetic and
observed integrated colours.\label{table:HRDs}}
  \begin{tabular}{@{}lccccc@{}}
  \hline
 & Isochrone & Age   & Synthetic & Observed$^{a}$ & \\
 & [Fe/H]    & (Gyr) & $(V-I)_0$ & $(V-I)_0$ & $\Delta (V-I)_0$\\
\hline
H10  & -1.31 & 12 & 0.94 & 0.95 & $-0.01$\\
 & & & & & \\
H23  & -1.31 & 9  & 0.98 & 1.01 & $-0.03$\\
 & & & & & \\
PA06 & -1.84 & 12 & 0.91 & 0.87 & $+0.04$\\
 & & & & & \\
PA17$^{b}$ & -1.01 & 12 & 1.05 & 1.14& $-0.09$\\
 & & & & & \\
PA53 & -1.62 & 12 & 0.89 & 0.85& $+0.04$\\
 & & & & & \\
PA54 & -1.84 & 13 & 0.89 & 0.84& $+0.05$\\
 & & & & & \\
PA56 & -1.62 & 12 & 0.88 & 0.89& $-0.01$\\
 & & & & & \\
\hline
\end{tabular}\\
\end{center}
\medskip
\raggedright $^{a}$ Uncertainties in the observed colours from
\citet{Huxor2014} are likely to be $\sim 0.05$ dex.\\
$^{b}$ PA17 does not yet have a high-quality {\it HST}
CMD; therefore, there are few {\it a priori} constraints on the
underlying stellar population.\\
\end{table}

\begin{table*}
\centering
\begin{minipage}{160mm}
\begin{center}
\caption{Trends in \ion{Fe}{1} \vspace{0.25in} abundance for the
PAndAS Clusters\label{table:PAndASSlopes}}
  \newcolumntype{d}[1]{D{,}{\pm}{#1} }
  \begin{tabular}{@{}ld{3}ccd{3}cd{3}c@{}}
 & & & \\
  \hline
Cluster & \multicolumn{3}{c}{\phantom{-----}Wavelength} &
\multicolumn{2}{c}{\phantom{---}REW  } & \multicolumn{1}{c}{\phantom{------}EP} &\\
        & \multicolumn{3}{c}{\phantom{-----}Slope     } &
\multicolumn{2}{c}{\phantom{---}Slope} & \multicolumn{1}{c}{\phantom{------}Slope} &\\
 & \multicolumn{3}{c}{\phantom{-----}{$10^{-5}$ dex/\AA}} & \multicolumn{2}{c}{\phantom{---}{(dex)}} & \multicolumn{1}{c}{\phantom{-----}{(dex/eV)}} \\
 & & & & & & & \\
 \hline
 & & & & & & & \\
H10  & \phantom{--}2.7,4.0 && & -0.030, 0.097 & & -0.011,0.019& \\
 & & & & & & & \\
H23  & \phantom{--}-1.5,2.9 && & -0.016,0.067  & & -0.0020,0.015& \\
 & & & & & & & \\
PA06 & \phantom{--}-1.8,2.3 && & -0.0057,0.058 & & 0.0041, 0.011& \\
 & & & & & & & \\
PA17$^{a}$ & \phantom{--}3.3,7.5 &&& 0.035, 0.19 & & 0.015, 0.029& \\
 & & & & & & & \\
PA53 & \phantom{--}3.7,5.7 && & 0.019, 0.18 &  & 0.019, 0.025& \\
 & & & & & & & \\
PA54 & \phantom{--}9.3,3.7 && & -0.00018,0.099& & -0.020,0.020& \\
 & & & & & & & \\
PA56 & \phantom{--}6.8,5.6 && & 0.033, 0.13   & & 0.0081,0.026& \\
 & & & & & & & \\
\hline
\end{tabular}
\end{center}
\end{minipage}\\
\medskip
\raggedright $^{a}$ PA17 does not yet have a high-quality {\it HST}
CMD; therefore, there are few {\it a priori} constraints on the
underlying stellar population.\\
\end{table*}

\subsection{Comparisons with Literature Values}\label{subsec:LitComparison}
\begin{figure*}
\begin{center}
\centering
\subfigure{\includegraphics[scale=0.65]{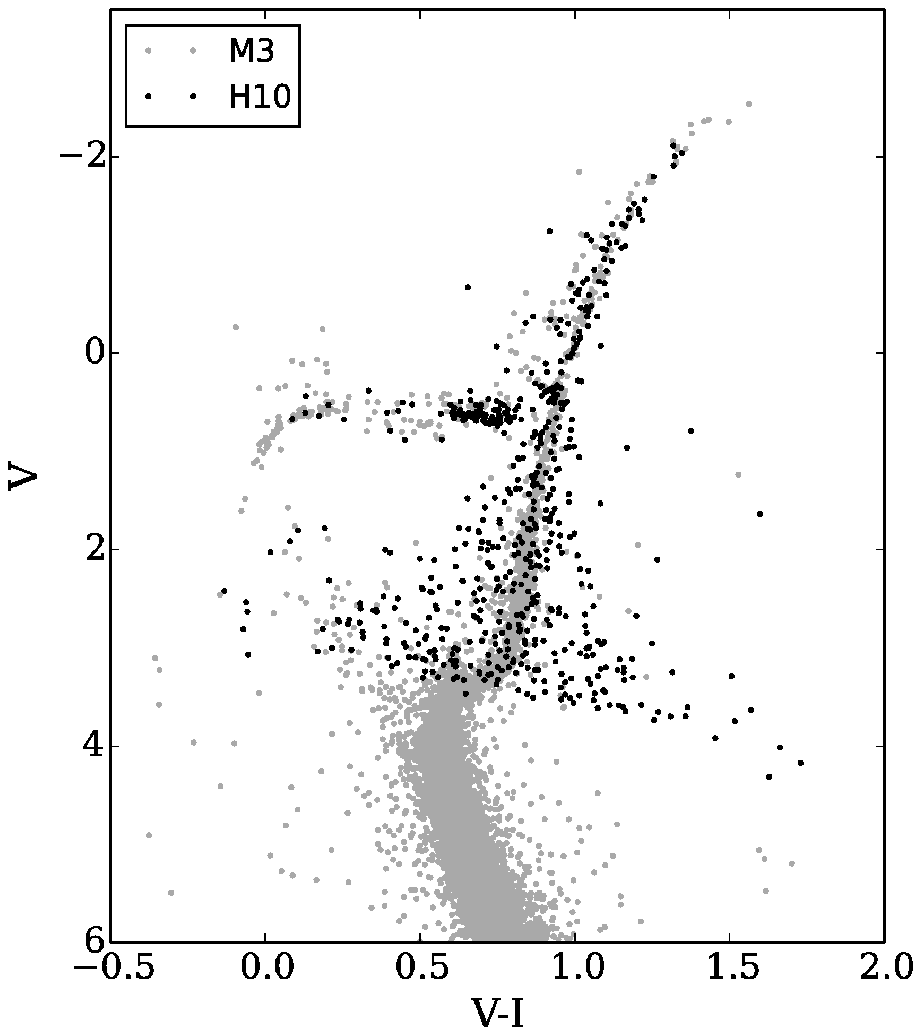}\label{fig:H10andM3}}
\subfigure{\includegraphics[scale=0.6]{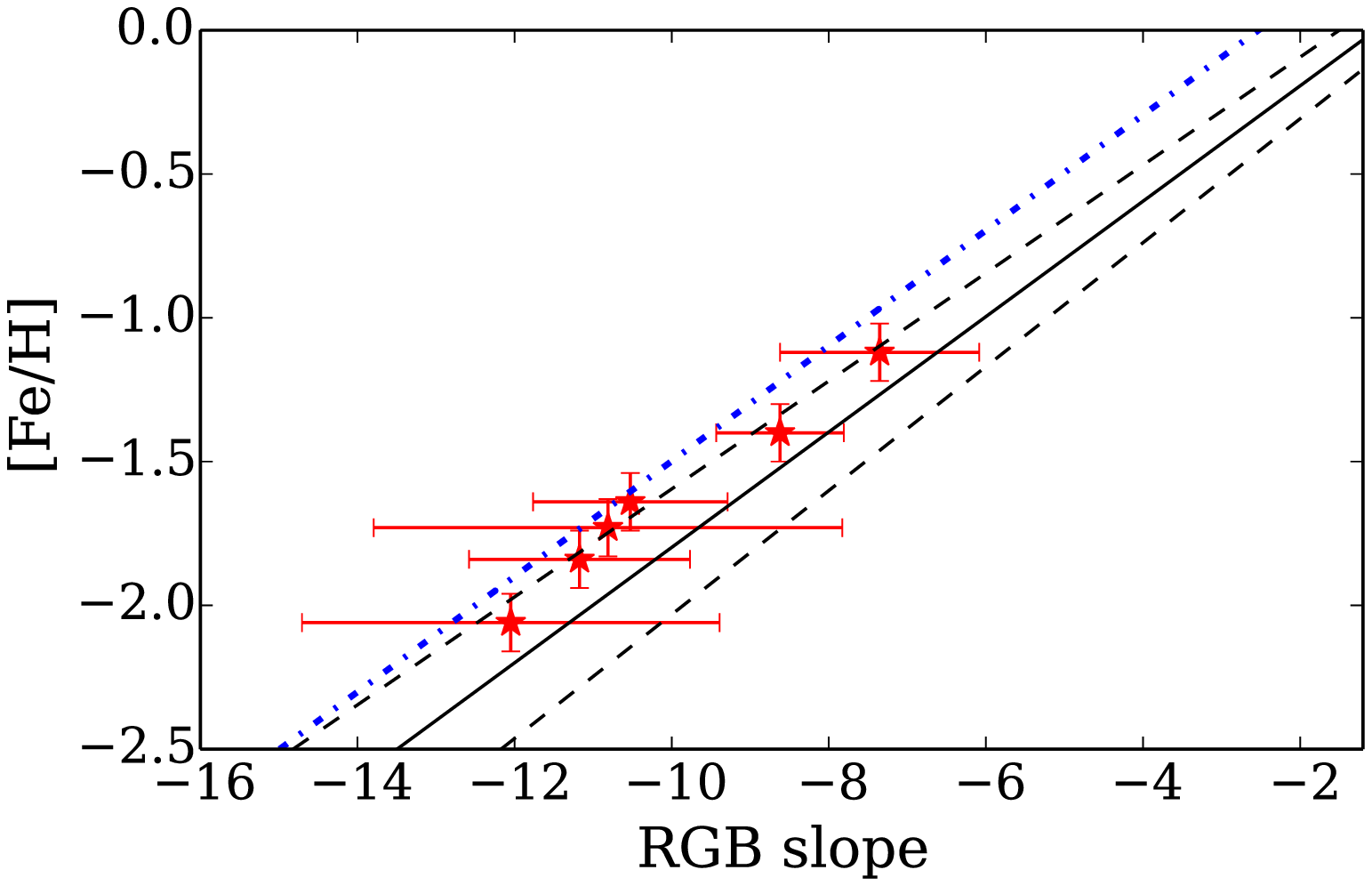}\label{fig:RGBslopes}}
\caption{{\it Left:} The H10 {\it HST} CMD, plotted on M3's $V$, $I$
CMD from the ACS Globular Cluster Survey
\citep{Sarajedini2007,Anderson2008}.  The distance modulus and
reddening have been adjusted to match the horizontal branches of the
two clusters.  The slopes of the GCs agree well, supporting the higher
spectroscopic metallicity from this study.
{\it Right:} RGB slopes (determined from the {\it HST} photometry)
versus spectroscopic metallicity.  The solid line shows the
calibration with Galactic GCs (presented in Appendix 
\ref{appendix:RGBslopes}), while the red stars show the slopes of the
PAndAS clusters.  The RGB slopes predict metallicities that agree well
with the integrated [Fe/H] values.  The blue dot-dashed line shows the
offset that would occur at a fixed [Fe/H] if clusters have
$[\alpha/\rm{Fe}] = 0$ (see Appendix
\ref{appendix:RGBslopes}).\label{fig:LitComparison}}
\end{center}
\end{figure*}

H10's isochrone age and integrated [Fe/H] agree well with the high
resolution IL values from \citet{Colucci2014}, which are determined
from spectra with a slightly different wavelength range.  However, the
metallicities and ages for H10 and H23 from this work do not agree
with the previous photometric results from \citet{Mackey2007}.  From
analyses of partially resolved CMDs of the upper RGBs,
\citet{Mackey2007} find [Fe/H] values for H10 and H23 that are lower
than the integrated [\ion{Fe}{1}/H] ratios presented here.  The
fiducial fits in Mackey et al. are performed with an optimization
routine that finds the best combination of Galactic GC fiducial (at a
given [Fe/H]), distance modulus, and reddening that fit the observed
RGB.  This should provide a good estimate of cluster [Fe/H], since the
slope of the RGB is primarily sensitive to metallicity
(e.g. \citealt{Sarajedini1994}), though this [Fe/H] is dependent on
the adopted distance modulus and reddening.  For instance,
\citet{Mackey2007} find that H10's RGB falls between the fiducials of
M92 ($[\rm{Fe/H}] = -2.14$) and NGC~6752 ($[\rm{Fe/H}] = -1.54$),
implying that H10 likely has a metallicity of $[\rm{Fe/H}] \approx
-1.84$.  However, Figure \ref{fig:H10andM3} demonstrates that H10's
RGB slope can be well represented by M3's RGB, indicating a higher
metallicity of $[\rm{Fe/H}] \approx -1.5$ (which agrees with the
spectroscopic values from this work).

A similar technique can be applied to all the PAndAS GCs.  An
empirical calibration between Galactic GC RGB slope (in the F606W and
F814W {\it HST} filters) is provided in Appendix
\ref{appendix:RGBslopes}. The RGB slopes and metallicities of the
PAndAS GCs are shown in Figure \ref{fig:RGBslopes} on top of this
empirical Galactic GC relation.  The cluster HBs were first aligned
with 47~Tuc and/or M3 to fix the distance modulus and reddening.  The
approximate $(V-I)_0$ colours at two fixed absolute magnitudes ($M_V =
0$ and $M_V = -2$) were then determined, and RGB slopes were
calculated.  The approximate uncertainties in RGB slope are estimated
from  the uncertainties in RGB colour and distance modulus---note that
PA56 has a large uncertainty because its paucity of bright HB stars
makes its distance modulus more uncertain.  The quoted errors in
[Fe/H] are 0.1 dex, to reflect possible systematic errors
\citep{Sakari2014}. The PAndAS GCs agree with the Galactic GC relation
within the $1\sigma$ errors; however, all the PAndAS GCs have slightly
steeper slopes than their Galactic counterparts, hinting that the
clusters may be $\alpha$-deficient compared to the calibrating MW GCs
(see the discussion in Appendix \ref{appendix:RGBslopes}). Regardless,
this general agreement shows that the spectroscopic metallicities
derived in this paper are consistent with the observed CMDs.

\section{Abundances}\label{sec:Abunds}
Abundances were determined with {\tt ILABUNDS} (first introduced in
\citealt{McWB}), an IL modification of the 1997 version of the Local
Thermodynamical Equilibrium (LTE) line analysis code {\tt MOOG}
\citep{Sneden}.  The final IL abundances\footnote{The standard
notation is used, i.e.
$$[\rm{X/H}] = \log \epsilon(\rm{X}) - \log \epsilon(\rm{X})_{\sun}
=\Big( 12+\log \frac{N_{\rm{X}}}{N_{\rm{H}}}\Big)  - \Big( 12+\log
\frac{N_{\rm{X}}}{N_{\rm{H}}}\Big)_{\sun}$$
where $N_{\rm{X}}$ is the column density of any element X.}
for all elements are shown in Table~\ref{table:Abunds}. Random errors
for EW-based abundances are calculated as in \citet{Shetrone2003} and
\citet{Sakari2011}, where the largest of three errors was adopted: the
line-to-line scatter for that element, the error due to EW
uncertainties, or the iron line-to-line scatter. The largest of these
three uncertainties is adopted as the final \textit{random} abundance
error for that element.  Random errors for abundances derived via
spectrum syntheses are calculated based on uncertainties in continuum
placement {\it and} line profile fitting (see \citealt{Sakari2013}).

Systematic errors are adopted from \citet{Sakari2014}, based on GC
[Fe/H] and HB morphology.  The optimal abundance ratios for chemical
tagging are those which
\begin{itemize}
\item are least sensitive to uncertainties in the underlying stellar
populations
\item are most distinct between massive and dwarf galaxies.
\end{itemize}
Sakari et al. determined that the most stable and useful abundance
ratios were [\ion{Ca}{1}/\ion{Fe}{1}], [\ion{Ni}{1}/\ion{Fe}{1}], and
[\ion{Ba}{2}/\ion{Eu}{2}].  The first two [X/Fe] ratios are typically
stable (within $\sim 0.1$ dex) to most systematic uncertainties in
the underlying populations.  Though individually
[\ion{Ba}{2}/\ion{Fe}{2}] and [\ion{Eu}{2}/\ion{Fe}{2}] are highly
sensitive to the underlying stellar populations, the \ion{Ba}{2} and
\ion{Eu}{2} offsets are often similar, and the [Ba/Eu] ratio is much
more stable to population uncertainties (in a partially resolved
system the uncertainties can still be as high as $\sim 0.1$ dex).
These ratios are also useful for comparisons with dwarf galaxy stars.
The [Ca/Fe] ratio can serve as an [$\alpha$/Fe] indicator, since metal
rich dwarf galaxy stars have lower [Ca/Fe] at a given [Fe/H] (see,
e.g., \citealt{Tolstoy2009}).  The [Ni/Fe] ratio has been observed to
be low in metal-rich field stars and clusters associated with the
Sagittarius (Sgr) dwarf spheroidal
\citep{Cohen2004,Sbordone2005,Sbordone2007} and anomalous MW field
stars that are suspected to have been accreted from dwarf galaxies
\citep{NissenSchuster2010}.  Finally, the [Ba/Eu] ratio is observed to
be high in dwarf galaxy stars (see, e.g.,
\citealt{Tolstoy2009}). Other suitable, high precision ratios include
[Eu/Ca] and [Mg/Ca]. Although the other element ratios
([\ion{Na}{1}/\ion{Fe}{1}], [\ion{Mg}{1}/\ion{Fe}{1}],
[\ion{Ti}{1}/\ion{Fe}{1}], and [\ion{Ti}{2}/\ion{Fe}{2}]) are less
stable to uncertainties in the underlying populations, their agreement
(or disagreement) with other abundance ratios may prove useful.

Table \ref{table:Sys} presents estimates of the maximum systematic
errors in the integrated abundances of the PAndAS clusters.  All
errors estimates are based on the tables presented in
\citet{Sakari2014}, with the exception of Na and Mg, which were
calculated in the same way.  The errors are selected from
\citet{Sakari2014} to match the metallicity and HB morphology of the
cluster, as well as the analysis technique.  The GCs H10, H23, PA06,
PA53, PA54, and PA56 are partially resolved, which reduces the
systematic errors considerably.  At $[\rm{Fe/H}]\sim -1.4$, H10 is
closest in [Fe/H] to M3, M13, and NGC~7006; its systematic
uncertainties are therefore averages from the three clusters (though
note that since all H10's HB stars are red, the hot star uncertainties
are not included). H23 has a metallicity in between 47~Tuc and
M3/M13/NGC~7006, and its quoted uncertainties are therefore an average
of the 47~Tuc values and the mean M3/M13/NGC~7006 values.  PA53, PA54,
and PA56 are similar in [Fe/H] and HB morphology to M3, M13, and
NGC~7006, and their averaged abundance offsets are therefore
utilized. M15's systematic uncertainties are adopted for PA06.  For
PA17, the unresolved cluster, the systematic uncertainties of 47~Tuc
were adopted (since PA17's $[\rm{Fe/H}] \sim -0.9$ is closest to
47~Tuc's $[\rm{Fe/H}] \sim -0.8$; see \citealt{Sakari2013}), assuming
that the cluster is modelled with isochrones and that there are no
constraints on the underlying population.

In the following subsections, each element type is discussed
separately.  The [Fe/H] and [X/Fe] ratios of the PAndAS GCs are
plotted along with field stars in the MW and its dwarf satellites.
There are no {\it detailed} abundances for M31 field stars, though
there are medium resolution $\alpha$-abundances for four outer halo
M31 stars \citep{Vargas2014}; however, M31 field star abundance
patterns are likely similar to the MW, especially for metal poor
stars.  The PAndAS clusters are also compared to integrated and
averaged abundances from GCs in the MW, the inner halo of M31, and MW
satellite galaxies.

\begin{table*}
\centering
\begin{minipage}{165mm}
\caption{Integrated abundances of the PAndAS clusters, with random
errors and the number of lines for each element.\label{table:Abunds}}
  \begin{tabular}{@{}lcccccccccc@{}}
  \hline
& PA17 & H23 & H10 & PA53 & PA56 & PA54 & PA06 \\
\hline
$[$\ion{Fe}{1}/H$]$ & $-0.93\pm0.03$ & $-1.12\pm0.02$ & $-1.36\pm0.02$ & $-1.64\pm0.03$ & $-1.73\pm0.03$ & $-1.84\pm0.02$ & $-2.06\pm0.02$ \\
$N$                 & 31 & 65 & 45 & 29 & 40 & 32 & 30 \\
 & & & & & & & \\
$[$\ion{Fe}{2}/H$]$ & $-0.96\pm0.15$ & $-1.21\pm0.08$ & $-1.30\pm0.04$ & $-1.68\pm0.10$ & $-1.71\pm0.10$ & $-1.75\pm0.08$ & $-2.09\pm0.10$\\
$N$                 & 4 & 5 & 5 & 1 & 1 & 2 & 1\\
 & & & & & & & \\
$[$\ion{Na}{1}/\ion{Fe}{1}$]$ & $0.60\pm0.15$& $-0.01\pm0.23$& $0.35\pm0.25$& $0.41\pm0.30$& $0.43\pm0.20$& $0.78\pm0.30$& $0.71\pm0.25$\\ 
$N$                 & 2 & 2 & 1 & 1 & 2 & 2 & 2 \\
 & & & & & & & \\
$[$\ion{Mg}{1}/\ion{Fe}{1}$]$ & $0.80\pm0.15$ & $0.30\pm0.20$ & $0.43\pm0.15$ & $0.20\pm0.20$ & $0.54\pm0.15$ & $0.32\pm0.18$ & $0.14\pm0.20$ \\ 
$N$                 & 1 & 2 & 2 & 1 & 2 & 2 & 1\\
 & & & & & & & \\
$[$\ion{Ca}{1}/\ion{Fe}{1}$]$ & $0.04\pm0.07$ & $0.41\pm0.04$ & $0.25\pm0.04$ & $0.19\pm0.03$ & $0.24\pm0.05$ & $0.28\pm0.04$ & $0.46\pm0.07$ \\ 
$N$                 & 4 & 6 & 10 & 10 & 8 & 9 & 9 \\
 & & & & & & & \\
$[$\ion{Ti}{1}/\ion{Fe}{1}$]$ & $0.0 \pm0.10$ & $0.56\pm0.05$ & $0.39\pm0.10$ & $0.16\pm0.10$ & $0.23\pm0.10$ & $-$ & $-$  \\
$N$                 & 1 & 3 & 1 & 1 & 1 & 0 & 0 \\
 & & & & & & & \\
$[$\ion{Ti}{2}/\ion{Fe}{2}$]$ & $0.17\pm0.10$ & $0.39\pm0.10$ & $0.40\pm0.03$ & $0.17\pm0.06$ & $-$ & $0.49\pm0.10$ & $0.31\pm0.10$ \\ 
$N$                 & 1 & 1 & 2 & 2 & 0 & 1 & 2 \\
 & & & & & & & \\
$[$\ion{Ni}{1}/\ion{Fe}{1}$]$ & $-0.08\pm0.09$ & $-0.07\pm0.04$ & $0.01\pm0.04$ & $-0.11\pm0.18$ & $-0.10\pm0.16$ & $-0.05\pm0.06$ & $-0.02\pm0.06$ \\
$N$                 & 3 & 5 & 7 & 2 & 2 & 3 & 2 \\
 & & & & & & & \\
$[$\ion{Ba}{2}/\ion{Fe}{2}$]$ & $0.21\pm0.25$ & $-0.03\pm0.11$ & $-0.05\pm0.11$ & $0.03\pm0.22$ & $-0.39\pm0.15$ & $-0.10\pm0.20$ & $-0.17\pm0.20$ \\ 
$N$                 & 2 & 2 & 2 & 2 & 1 & 1 & 1\\
 & & & & & & & \\
$[$\ion{Eu}{2}/\ion{Fe}{2}$]$ & $0.36\pm0.35$ & $0.53\pm0.25$ & $0.75\pm0.20$ & $0.68\pm0.25$ & $0.96\pm0.15$ & $<0.50$ & $<1.04$ \\
$N$                 & 1 & 1 & 1 & 1 & 1 & 1 & 1 \\
 & & & & & & & \\
\hline
\end{tabular}
\end{minipage}\\
\medskip
\raggedright {\bf Notes: } [Fe/H] and [X/Fe] ratios are calculated
differently, {\it line by line}, relative to the solar abundances
for that line (see \citealt{Sakari2013} and \citealt{Sakari2014} for
the EWs used to derive the solar abundances).\\
\end{table*}

\begin{table*}
\centering
\begin{minipage}{165mm}
\begin{center}
\caption{Systematic error estimates.\label{table:Sys}}
  \begin{tabular}{@{}lcccccccccccccc@{}}
  \hline
 & \multicolumn{2}{l}{$ \Big | \Delta[$Fe/H$] \Big | $} & \multicolumn{8}{l}{$ \Big | \Delta[$X/Fe$] \Big | $} & \multicolumn{4}{l}{$ \Big | \Delta[$X/Y$] \Big |$} \\
& \ion{Fe}{1} & \ion{Fe}{2} & \ion{Na}{1} & \ion{Mg}{1} & \ion{Ca}{1}
& \ion{Ti}{1} & \ion{Ti}{2} & \ion{Ni}{1} & \ion{Ba}{2} & \ion{Eu}{2}
& [Mg/Ca] & [Na/Mg] & [Ba/Eu] & [Eu/Ca]\\
\hline
{\bf PA17}             & 0.19 & 0.21 & 0.21 & 0.27 & 0.13 & 0.16 & 0.17 & 0.07 & 0.23 & 0.13 & 0.16 & 0.15 & 0.13 & 0.11 \\
{\bf H23}              & 0.13 & 0.19 & 0.13 & 0.14 & 0.08 & 0.12 & 0.10 & 0.07 & 0.18 & 0.10 & 0.08 & 0.11 & 0.09 & 0.09 \\
{\bf H10 }             & 0.09 & 0.15 & 0.11 & 0.07 & 0.05 & 0.08 & 0.08 & 0.07 & 0.09 & 0.08 & 0.04 & 0.10 & 0.07 & 0.10 \\
{\bf PA53, PA54, PA56} & 0.11 & 0.16 & 0.13 & 0.08 & 0.05 & 0.08 & 0.11 & 0.08 & 0.10 & 0.09 & 0.05 & 0.11 & 0.11 & 0.10 \\
{\bf PA06}             & 0.20 & 0.13 & 0.10 & 0.08 & 0.07 & 0.11 & 0.09 & 0.07 & 0.16 & 0.10 & 0.07 & 0.10 & 0.12 & 0.12 \\
 & & & & & & & & & & & & \\
\hline
\end{tabular}
\end{center}
\end{minipage}\\
\medskip
\raggedright {\bf Notes: } Total errors are conservatively estimated
by adding the individual errors from \citet{Sakari2014} in quadrature,
based on the appropriate [Fe/H] and HB morphology.\\
\end{table*}

\subsection{Iron}\label{subsec:PAndASFe}
The [Fe/H] abundances were determined with EWs, as described in
Section \ref{subsec:DAOSPEC}.  Because 1) there are fewer \ion{Fe}{2}
lines and 2) \ion{Fe}{2} is more susceptible to systematic effects
\citep{Sakari2013}, [\ion{Fe}{1}/H] is adopted as the representative
[Fe/H], despite potential non-LTE (NLTE) effects.  Note that for all
clusters \ion{Fe}{1} and \ion{Fe}{2} are in reasonably good agreement.
Three clusters, H10, H23, and PA17, are more metal-rich than
$[\rm{Fe/H}]~=~-1.5$, while the other four are metal-poor.  The most
metal-poor clusters all have fewer detectable lines ($\sim 30$
\ion{Fe}{1} lines and 1-2 \ion{Fe}{2} lines) than the more metal-rich
GCs, leading to larger random errors in [Fe/H].

Most GC systems in dwarf galaxies tend to be metal-poor, with more
massive dwarf galaxies possessing more metal-rich GCs
(e.g. \citealt{Peng2006}).   The comparatively high metallicities of
H10, H23, and PA17 therefore suggest that they formed in a galaxy more
massive than the Fornax (For) dwarf spheroidal.  It is clear from
Tables~\ref{table:PAndASTargets} and \ref{table:Abunds} that the
metal-rich PAndAS GCs have smaller projected distances from the centre
of M31 than the more metal-poor GCs. Based on a low resolution survey
of inner halo, bulge, and disk M31 GCs, \citet{Caldwell2011} found
evidence for an abundance gradient in the inner regions of M31---this
gradient seems to flatten to $[\rm{Fe/H}] \sim -1.8$ at distances $\ga
2$ kpc. H23 and PA17 are therefore more metal-rich than expected given
their large distance from the centre of M31; however, metal-rich field
stars {\it have} been identified in streams in M31's outer halo
\citep{Ibata2014}.

\subsection{$\alpha$-elements}\label{subsec:PAndASAlpha}
The $\alpha$-elements (Mg, Ca, and Ti) form primarily through captures
of $^{4}$He nuclei during hydrostatic burning in massive stars, while
Fe forms both during core collapse supernovae of massive stars {\it
and} during the detonation of white dwarfs whose progenitors were
lower mass stars.  Comparisons between the abundances of
$\alpha$-elements and Fe vs. [Fe/H] provide an indication of how the
chemical contributions from different types of stars have changed in a
specific environment.  Because the trends in [$\alpha$/Fe] with [Fe/H]
differ between massive and dwarf galaxies, the abundance ratios can
be used to link stars and/or GCs to their birth environments.

\subsubsection{Magnesium}\label{subsubsec:Mg}
\begin{figure*}
\begin{center}
\centering
\subfigure{\includegraphics[scale=1.0]{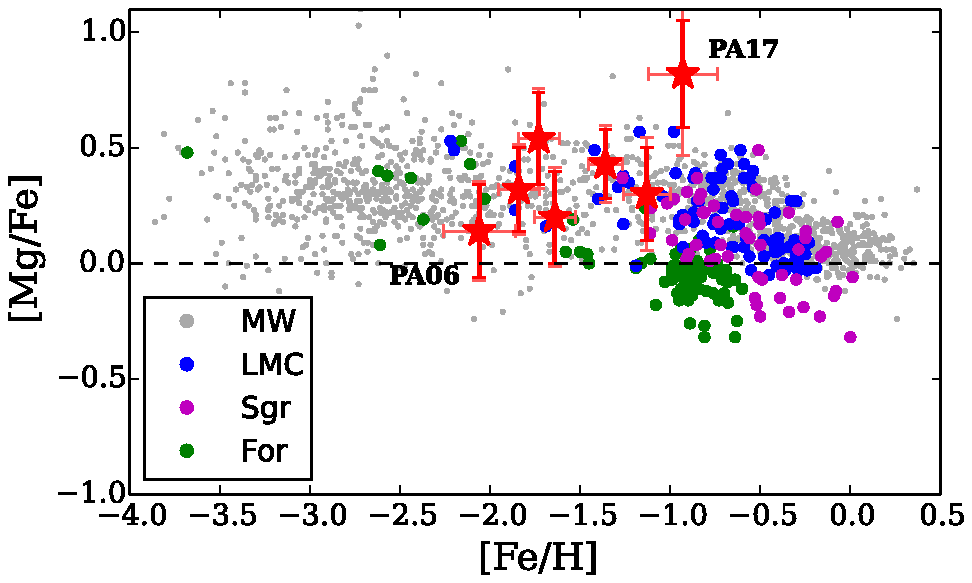}\label{fig:PAndASMgFSs}}
\subfigure{\includegraphics[scale=1.0]{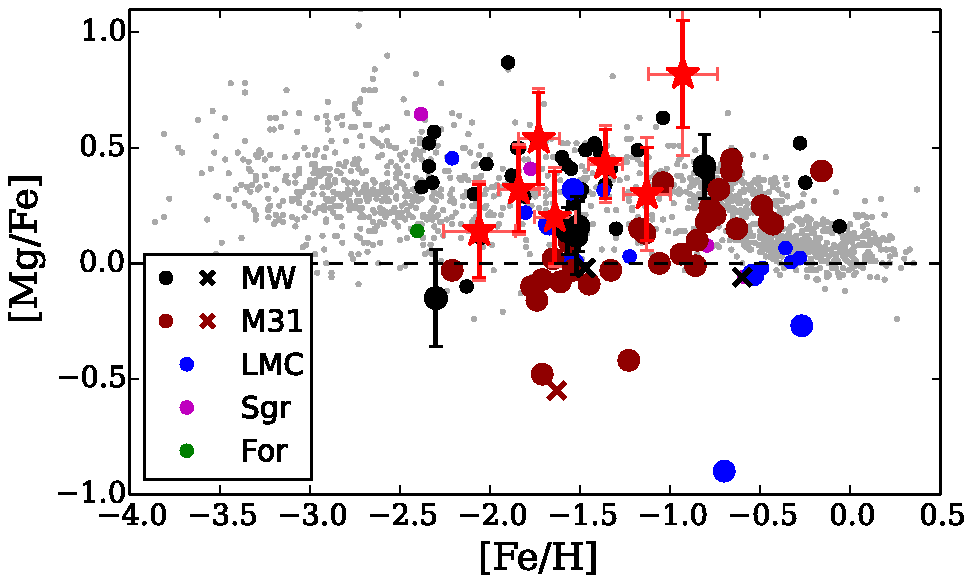}\label{fig:PAndASMgGCs}}
\caption{{\it Top: } Comparisons of PAndAS clusters (red stars) to
MW field stars (grey; from the sources in \citealt{Venn2004}, with
supplements from \citealt{Reddy2006}) and dwarf galaxy field stars.
LMC stars are shown in blue \citep{Pompeia}, Sgr in magenta
\citep{Sbordone2007,Monaco2007,CarrettaSgr,ChouSgr,McW2013}, and
For in green \citep{Tafelmeyer2010,Shetrone2003,Letarte2010}.  The
error bars show the random errors.  The dashed horizontal line shows
the solar value.  Random errors are shown as thick error bars, while
systematic and random errors (added in quadrature) are shown with the
thinner error bars.
{\it Bottom: } Comparisons of PAndAS clusters (red stars) to MW field
stars (grey), MW clusters (black circles), other M31 clusters
(maroon), and dwarf galaxy GCs.  IL abundances are shown as larger
symbols, while averaged GC abundances from individual stars are shown
as smaller symbols.  IL abundances of MW clusters are from
\citet{Sakari2014}, while cluster averages are from
\citet{Pritzl2005}.  The M31 cluster IL abundances are from
\citet{Colucci2009,Colucci2014}.  LMC clusters are in blue; IL
abundances are from \citet{Colucci2012}, while the individual stellar
abundances are from \citet{Johnson2006} and \citet{Mucciarelli2008}.
Average abundances of Sgr clusters are from \citet{Cohen2004},
\citet{Sbordone2005}, and \citet{Mottini2008}. Averages of the
individual stars in For clusters are in green, and are from
\citet{Letarte2006}.  The unusual MW clusters Palomar~1 and Rup 106
(from \citealt{Sakari2011} and \citealt{Villanova2013}) and the M31 GC
G002 (from \citealt{Colucci2014}) are shown with crosses---these GCs
may have been accreted from dwarf galaxies, though they have no
obvious host galaxies.\label{fig:PAndASMg}}
\end{center}
\end{figure*}

Magnesium abundances for the PAndAS GCs are presented in Table
\ref{table:Abunds}.  For most clusters [Mg/Fe] is determined from the
5528 and 5711 \AA \hspace{0.025in} lines---however, PA17's [Mg/Fe] is
only determined from the 5711 \AA \hspace{0.025in} line because its
5528 \AA \hspace{0.025in} line is too strong; similarly, the 5711 \AA
\hspace{0.025in} line is too weak in the most metal-poor cluster,
PA06, and in the GC with the highest velocity dispersion, PA53.
Comparisons with MW, M31, and dwarf galaxy field stars and clusters
are shown in Figure \ref{fig:PAndASMg}. Field stars in the most
massive, nearby, well-studied dwarfs that host GCs are also shown,
including the LMC and the Sgr and For dwarf spheroidals; lower mass
galaxies such as Sculptor and Carina are not included since they do
not have GCs.  The average and/or integrated abundances of MW, M31,
LMC, Sgr, and For GCs are shown separately from the field stars.
The unusual MW GCs Palomar~1 (Pal~1; \citealt{Sakari2011}) and
Ruprecht 106 (Rup 106; \citealt{Villanova2013}) and the M31 GC
G002 \citet{Colucci2014} are given different symbols, since their
chemical abundances indicate that they may have originated in dwarf
galaxies (though their host galaxies have not yet been identified).

Note that star-to-star Mg variations {\it have} been observed in
massive, metal-poor Galactic clusters ($[\rm{Fe/H}] \la -1.2$;
\citealt{Carretta2009}); in those clusters, Mg variations can be as
large as $0.5-1$ dex (e.g. \citealt{Sneden1997,Sneden2004},
\citealt{CohenMelendez2005}). The most metal-poor PAndAS GC, PA06, has
a low [Mg/Fe] ratio for its metallicity, similar to M15 (see
\citealt{Sakari2013}); this suggests that strong star-to-star Mg
variations may exist in PA06. The integrated [Mg/Fe] ratios in the
other metal-poor GCs roughly agree with the Galactic GCs.  The
integrated Mg abundances of the \citet{Colucci2014} M31 GCs are
systematically lower than the PAndAS GCs at the metal poor
end---however, this may be due to the masses of the observed
clusters.  Many of the Colucci et al. targets have larger velocity
dispersions and higher total luminosities, indicating that they are
more massive than these PAndAS GCs.  If this is the case, the GCs in
the Colucci et al. sample may harbour stronger Mg variations
(e.g. \citealt{Carretta2009}). \citet{Colucci2014} {\it do} find a
correlation between integrated [Mg/Fe] and total $M_V$ for the metal
poor GCs, such that the more massive GCs have lower integrated
[Mg/Fe]---all five of the most metal poor ($[\rm{Fe/H}] < -1.2$)
PAndAS GCs are consistent with the Colucci et al. relation for all
GCs.

\begin{figure*}
\begin{center}
\centering
\includegraphics[scale=0.7]{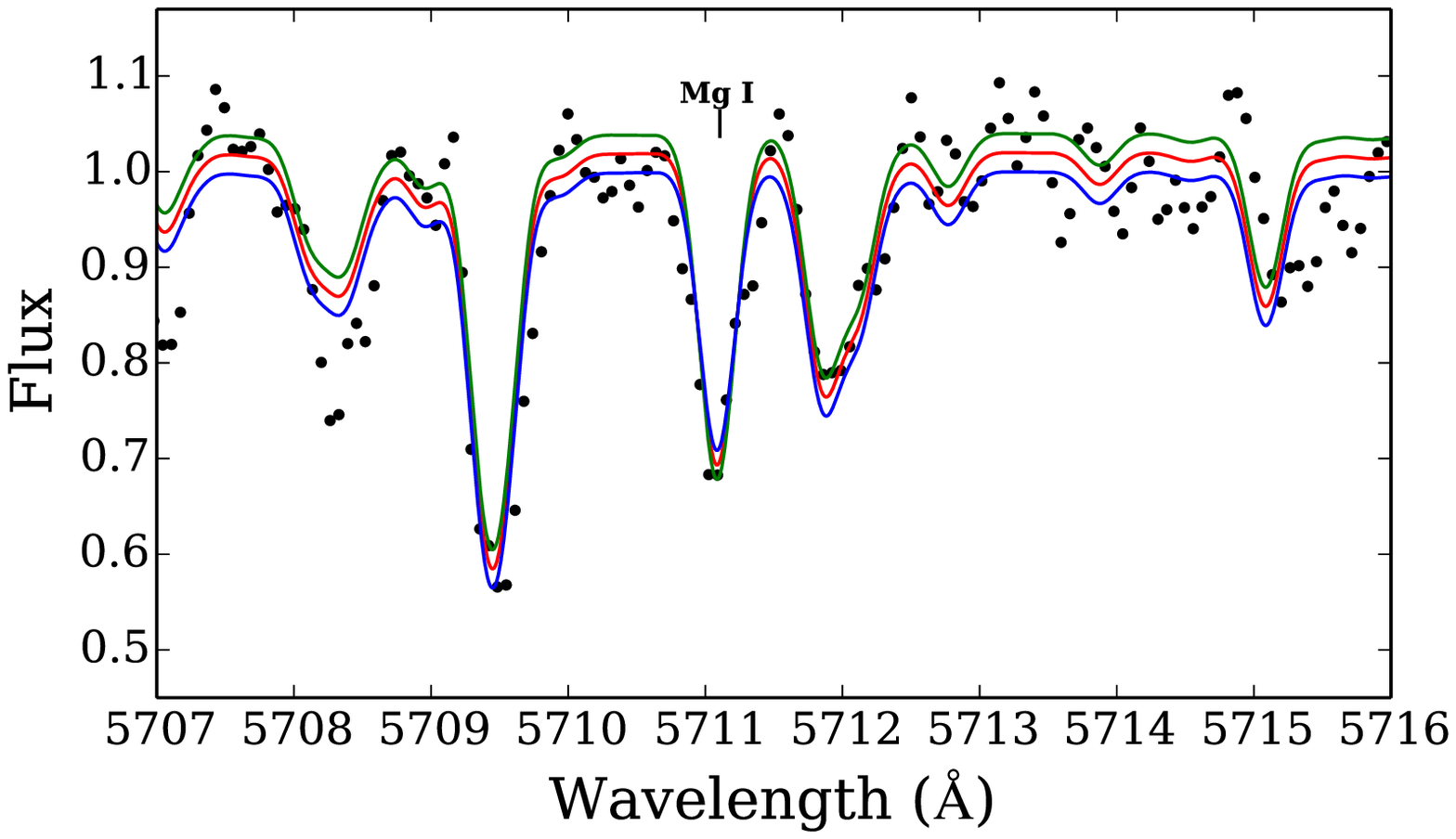}
\caption{Syntheses of the 5711 \AA \hspace{0.025in} \ion{Mg}{1} line
in PA17.  The red line shows the best fit, while the green and blue
lines are the $\pm 1\sigma$ fits.}\label{fig:SynthMg}
\end{center}
\end{figure*}

Because of their high metallicities the two most metal-rich PAndAS
GCs, H23 and PA17, are not likely to host significant star-to-star
variations in Mg, though they may host multiple populations (see
Section \ref{subsubsec:NaMg}).  The [Mg/Fe] ratios of H10 and H23
agree with Galactic field stars.  Note that H10's [Mg/Fe] is higher
than the value presented by \citet{Colucci2014}, which was derived
from slightly different lines; considering random and systematic
uncertainties, however, the two are consistent.  PA17 has a [Mg/Fe]
ratio of 0.8 dex, which is higher than most of the field stars and all
of the GCs.  The syntheses of PA17's 5711 \AA \hspace{0.025in} line
are shown in Figure \ref{fig:SynthMg}---this line has a similar
strength in all 3 exposures, suggesting that the line strength is not
due to an improperly removed cosmic ray. Note that there are no other
\ion{Mg}{1} features in the observed spectrum (other than the strong
5528 \AA \hspace{0.025in} line) to verify this high abundance.  With
its high [Mg/Fe], PA17 falls at the upper end of the Galactic and LMC
field stars.  This high [Mg/Fe] ratio suggests a greater contribution
from massive star ejecta, either from a top-heavy IMF or from
inhomogeneous mixing and stochastic sampling of supernova
ejecta. Alternatively, high Mg could indicate contributions from a
rapidly rotating massive star that went supernova
(e.g. \citealt{Takahashi2014}).

\subsubsection{Calcium and Titanium}\label{subsubsec:CaTi}
Calcium and titanium mainly form explosively in core collapse
supernovae and generally behave like the other $\alpha$-elements,
though both elements also form in Type Ia supernovae
\citep{Nomoto1984,Maeda2010}.  The integrated
[\ion{Ca}{1}/\ion{Fe}{1}] abundances are  determined based on EWs of
$\la 10$ spectral lines. All \ion{Ca}{1} line EWs were verified by
hand, and any uncertain lines were removed from the analysis. Spectrum
syntheses were also performed on several \ion{Ca}{1} lines to verify
the EW-based abundances.  The Ti abundances are based on fewer lines
(1-3), but both [\ion{Ti}{1}/\ion{Fe}{1}] and
[\ion{Ti}{2}/\ion{Fe}{2}] generally agree with [Ca/Fe].
\citet{Sakari2014} determined that the [Ti/Fe] ratios are more
sensitive to systematic uncertainties in the underlying stellar
populations; thus, only [Ca/Fe] is compared with other targets.

\begin{figure*}
\begin{center}
\centering
\subfigure{\includegraphics[scale=1.0]{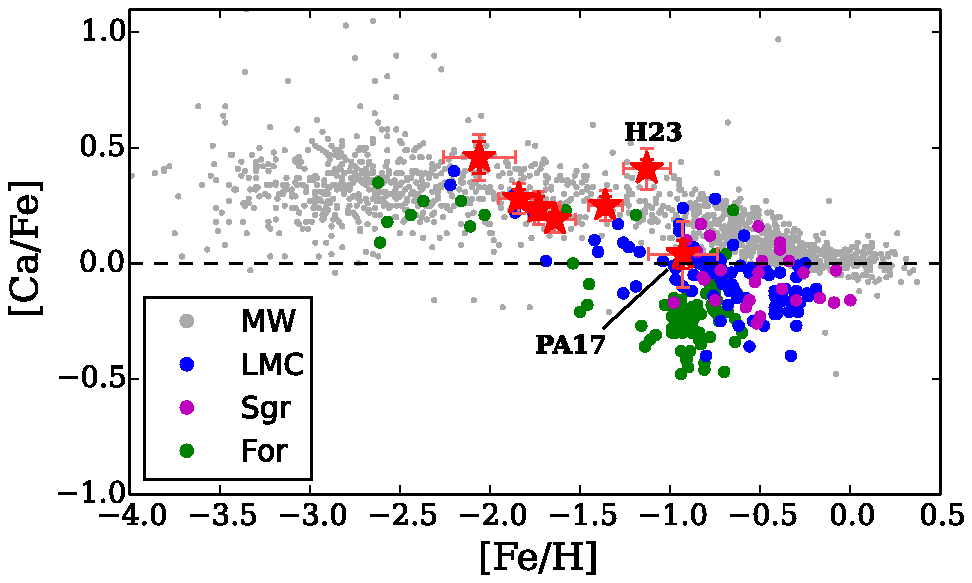}\label{fig:PAndASCaFSs}}
\subfigure{\includegraphics[scale=1.0]{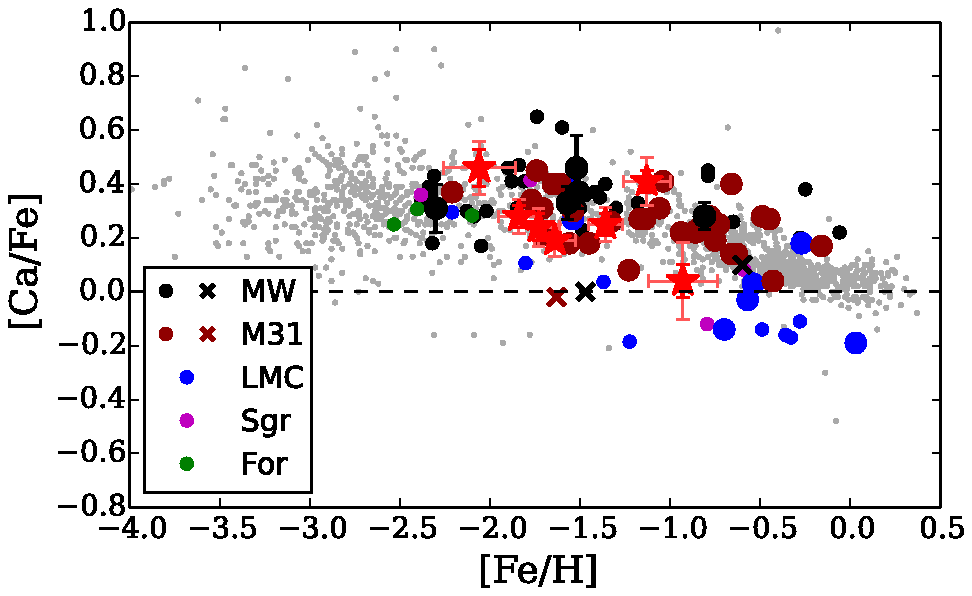}\label{fig:PAndASCaGCs}}
\caption{{\it Top: } Comparisons of PAndAS clusters (red stars) to
MW field stars and dwarf galaxy field stars. Points and references are
as in Figure \ref{fig:PAndASMgFSs}.
{\it Bottom: } Comparisons of PAndAS clusters (red stars) to MW field
stars and GCs from the MW, M31, and various dwarf galaxies.  Points
and references are as in Figure
\ref{fig:PAndASMgGCs}.\label{fig:PAndASCa}}
\end{center}
\end{figure*}

The PAndAS cluster [Ca/Fe] ratios are compared to MW and dwarf galaxy
field stars in Figure \ref{fig:PAndASCaFSs} and to MW, M31, and dwarf
galaxy GCs in Figure \ref{fig:PAndASCaGCs}.  Most of the PAndAS GCs
are Ca-enhanced, though PA17's [Ca/Fe] is roughly solar.  All the
PAndAS clusters look very similar to the LMC and Sgr GCs; most of the
For clusters are more metal-poor than the PAndAS GCs.  H10's [Ca/Fe]
agrees well with \citet{Colucci2014} and is slightly lower than the
average MW and M31 clusters.  PA17's low [Ca/Fe] could be understood
several ways.  As a $\alpha$-element, low [Ca/Fe] could be indicative
of the slow star formation ratios or top-light IMFs seen in dwarf
galaxies (e.g. \citealt{Tinsley1979,McW2013})---however, PA17's high
[Mg/Fe] ratio suggests that this interpretation is not so
simple. Regardless of the cause, PA17's Ca abundance looks most
similar to Pal~1, Ter~7, Pal~12, and the LMC field stars and
clusters.

\subsubsection{Mg vs. Ca}\label{subsubsec:MgCa}
If Mg and Ca have the same nucleosynthetic site then the [Mg/Ca]
ratio should not change with [Fe/H].  In Galactic field stars this
is more or less what is observed: [Mg/Ca] is roughly solar except at
the lowest metallicities where the dispersion is very high.  The
dispersion at low metallicities is typically ascribed to inhomogeneous
mixing (see, e.g., \citealt{Venn2012}).  However, metal-rich dwarf
galaxy stars appear to have different [Mg/Ca] ratios from MW stars.
In fact, the nucleosynthetic sites of Mg and Ca are {\it not} the
same: Mg is mainly produced during the hydrostatic burning of massive
stars  (e.g. \citealt{WoosleyWeaver1995}) while Ca is formed
explosively in both Type II and Ia supernovae
(e.g. \citealt{WoosleyWeaver1995}, \citealt{Nomoto1984},
\citealt{Maeda2010}).  Figure \ref{fig:PAndASMgCa} shows that the
metal-rich field stars in the LMC and For have higher [Mg/Ca]
ratios than MW stars at the same metallicity---this discrepancy is
typically ascribed to the different nucleosynthetic sites of Mg and Ca
(e.g. \citealt{Shetrone2003,Venn2004,Letarte2010,Venn2012}).  However,
stars in Sgr have low [Mg/Ca] ratios (except possibly for stars in the
Sgr streams; \citealt{Monaco2007}); \citet{McW2013} attribute these
low ratios to a top-light IMF, where the lack of massive stars means
that less Mg is produced than Ca.   Most of the PAndAS GCs overlap
with Galactic and dwarf galaxy stars; however, PA17's [Mg/Ca] ratio is
higher than the Galactic stars, in agreement with the LMC stars.

Again, star-to-star Mg variations could affect the integrated [Mg/Ca]
ratios of the metal-poor GCs, complicating comparisons with field
stars.  However, if the integrated [Mg/Fe] ratios are affected by
multiple populations, one might expect the integrated [Mg/Fe] and
[Mg/Ca] ratios to be {\it lower} than the ``primordial'' values
(because Mg is expected to decrease in the second population). Only
PA06 has a lower [Mg/Ca] than the majority of the Galactic stars, in
agreement with the integrated [Mg/Ca] of M15, which has known, {\it
strong} star-to-star Mg variations that affect the integrated [Mg/Fe]
\citep{Sneden1997,Sakari2013}.  Thus, PA06 seems to be similar to M15,
and may host a significant second, Mg-deficient population.  The other
metal-poor GCs agree well with the Galactic stars and GCs, with the
exception of PA56, which has a slightly high [Mg/Ca].  Note that most
of the PAndAS GCs have higher [Mg/Ca] ratios than the M31 GCs---again,
this is likely because the GCs in the \citet{Colucci2014} sample are
more massive than the PAndAS GCs, and therefore may have stronger
star-to-star Mg variations.

The GCs more metal rich than $[\rm{Fe/H}] \sim -1.2$ are not likely to
host significant Mg variations.  H10 and H23 have roughly solar
[Mg/Ca] ratios, in agreement with the Galactic field stars and
clusters (though \citealt{Colucci2014} find a lower [Mg/Ca] for H10
because of their lower [Mg/Fe]).  PA17 is higher than the Galactic
field stars and clusters by $\sim 2\sigma$, in agreement with the LMC
and Fornax field stars.  This high [Mg/Ca] is driven by the high
[Mg/Fe] (determined from spectrum syntheses of the 5711 \AA
\hspace{0.025in} line) and the low [Ca/Fe] (from EWs of four clean
lines).  Again, the high Mg suggests that PA17 is enriched with the
products of the most massive stars, which may be due to IMF effects,
stochastic sampling of supernova ejecta, or rotating supernova
progenitors (\citealt{Takahashi2014} also note that rotating massive
stars will not produce as much Ca, compared to non-rotating stars).
Despite its agreement with the LMC field stars, PA17 is distinct from
the LMC GCs, because the [Mg/Ca] ratios of the metal rich LMC GCs are
not high like the LMC field stars.   For the GC with the lowest
[Mg/Ca] ratio (NGC~1718), \citet{Colucci2012} suggest an inhomogeneous
mixing scenario, where NGC~1718 formed out of material with no
contributions from the highest mass stars.  However, the three metal
rich, [Mg/Ca]-poor LMC GCs have intermediate ages (1-3 Gyr), and it
may not be entirely appropriate to compare these intermediate-age GCs
with the older PAndAS GCs.  Regardless of the disagreement with the
LMC GCs, PA17's high [Mg/Ca] ratio is most similar to the LMC and
Fornax field stars.

\begin{figure*}
\begin{center}
\centering
\subfigure{\includegraphics[scale=1.0]{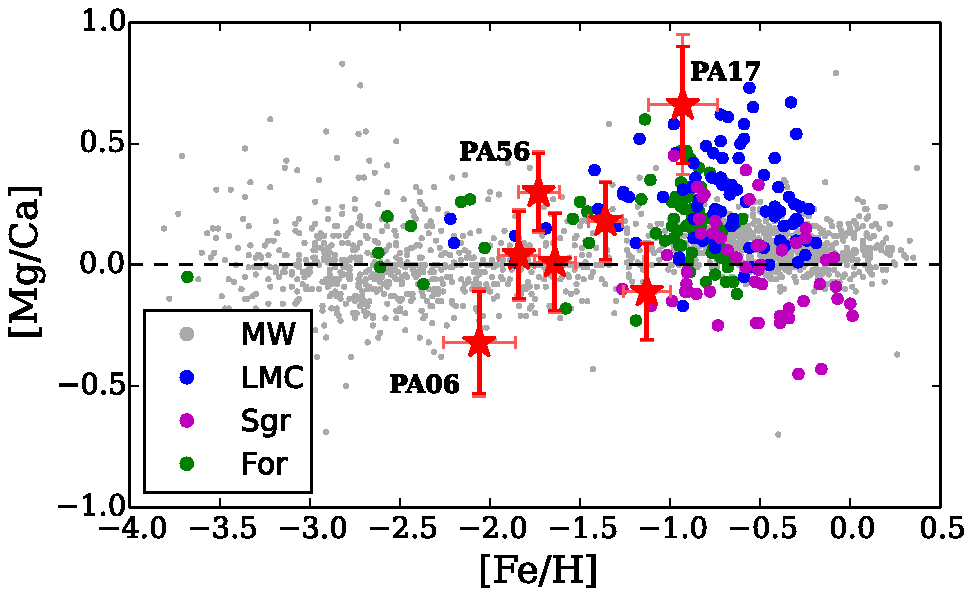}\label{fig:PAndASMgCa}}
\subfigure{\includegraphics[scale=1.0]{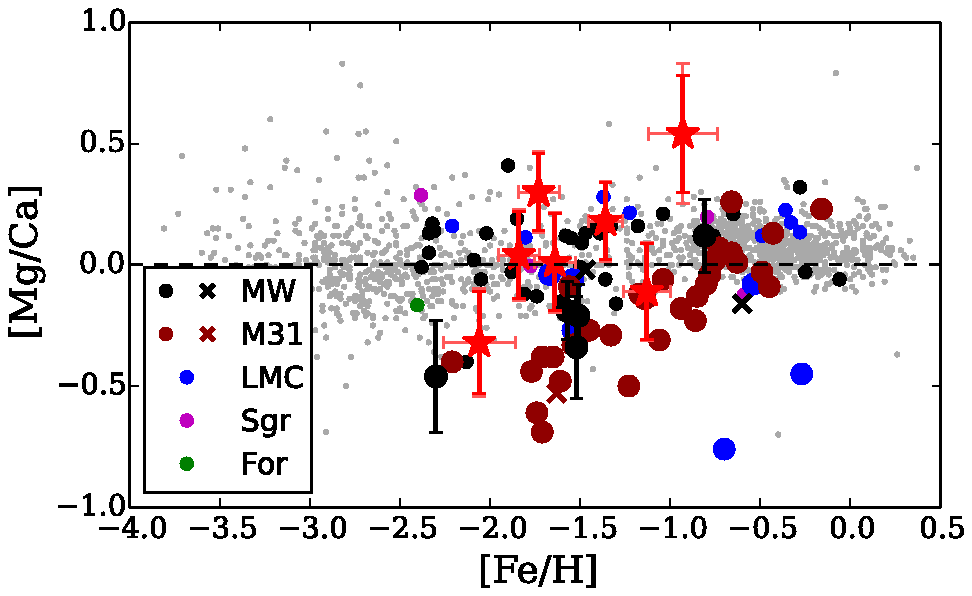}\label{fig:PAndASMgCaGCs}}
\caption{{\it Top: } Comparisons of PAndAS clusters (red stars) to
MW field stars and dwarf galaxy field stars. Points and references are
as in Figure \ref{fig:PAndASMgFSs}.
{\it Bottom: } Comparisons of PAndAS clusters (red stars) to MW field
stars and GCs from the MW, M31, and various dwarf galaxies.  Points
and references are as in Figure
\ref{fig:PAndASMgGCs}.\label{fig:PAndASMgCa}}
\end{center}
\end{figure*}


\subsection{Sodium}\label{subsec:PAndASLight}
Sodium abundances are derived from spectrum syntheses of the 6154 and
6160 \AA \hspace{0.025in} lines, which are all weaker than 100 m\AA
\hspace{0.025in} in the PAndAS IL spectra.  Sodium lines can be
particularly sensitive to NLTE effects; however, the 6154/6160 \AA
\hspace{0.025in} lines are not expected to have significant NLTE
corrections ($<0.2$ dex) at subsolar metallicities and for weak line
strengths (e.g. \citealt{Lind2011}).\footnote{Though note that some of
the HRD boxes (see Section \ref{sec:ModelAtms}) can have EWs $\sim
130$ m\AA \hspace{0.025in} in the most metal-rich clusters.} No
corrections were applied to the [Na/Fe] abundances, which are shown in
Table \ref{table:Abunds}.

\subsubsection{[Na/Fe]}\label{subsubsec:NaFe}
\begin{figure*}
\begin{center}
\centering
\subfigure{\includegraphics[scale=1.0]{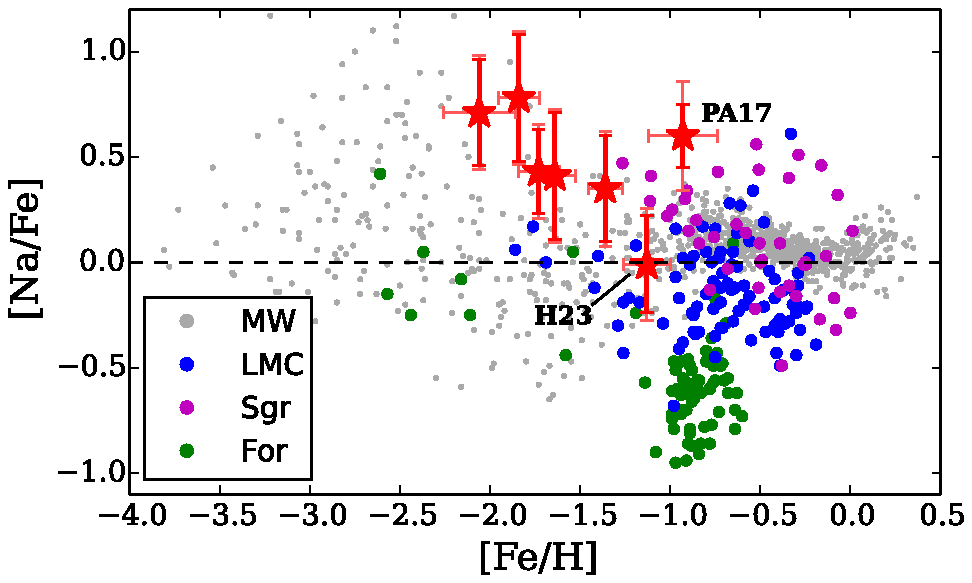}\label{fig:PAndASNaFSs}}
\subfigure{\includegraphics[scale=1.0]{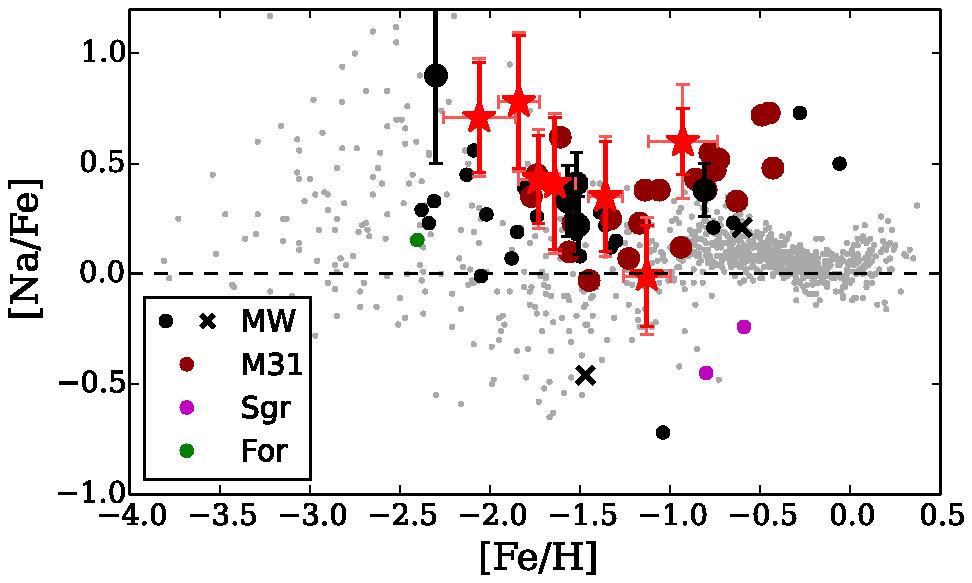}\label{fig:PAndASNaGCs}}
\caption{{\it Top: } Comparisons of PAndAS clusters (red stars) to
MW field stars and dwarf galaxy field stars.  Points and references
are as in Figure \ref{fig:PAndASMgFSs}.
{\it Bottom: } Comparisons of PAndAS clusters (red stars) to MW field
stars and GCs from the MW, M31, and various dwarf galaxies.  Points
and references are as in Figure
\ref{fig:PAndASMgGCs}.\label{fig:PAndASNa}}
\end{center}
\end{figure*}

Figure \ref{fig:PAndASNa} shows comparisons between the PAndAS GCs and
the field stars and GCs associated with the various galaxies.  Several
of the PAndAS GCs have [Na/Fe] ratios that are higher than MW and
dwarf galaxy field stars.  PA06, PA17, and PA54 all have high [Na/Fe]
ratios ($>~0.6$ dex) which places them above the Galactic and dwarf
galaxy stars, while the H10, PA53 and PA56 ratios are mildly high
($\sim 0.4$ dex).  These high values agree well with the IL values of
the Galactic and old LMC GCs and with many of the averaged Galactic
GCs.  In Galactic GCs the [Na/Fe] ratio is affected by star-to-star
abundance variations within the clusters, which are often  observed as
Na/O anticorrelations (see \citealt{Carretta2009}).  These abundance
variations are typically interpreted as arising from two chemically
distinct stellar populations within the GCs, which may indicate two
separate {\it generations} of stars (see the review by
\citealt{Gratton2012}). Several Galactic GCs have more centrally
concentrated Na-enhanced, O-deficient populations (e.g. M13 and
47~Tuc; \citealt{JohnsonPilachowski2012,Cordero2014}).  IL spectra of
the central regions (such as the Galactic GCs in
\citealt{Sakari2013,Sakari2014}) will likely be dominated by the
second generation (Na-enhanced, O-deficient) populations.  It is
therefore probable that (like the Galactic GCs) the PAndAS GCs are
Na-enhanced because of the presence of multiple populations.  It is
possible that H23's solar [Na/Fe] is due to the larger coverage of the
IL spectrum (see Table \ref{table:PAndASObservations}), which may
include light from the less centrally concentrated first generation
stars; alternatively, H23 may have more ``primordial'' stars.  Again,
the \citet{Colucci2014} [Na/Fe] ratio for H10 is lower than the one
derived here---however, the ratios are derived from only two lines in
both cases and are consistent within random and systematic errors.

Thus, the [Na/Fe] ratios indicate that these PAndAS GCs are likely to
be ``classical'' GCs under the definition of \citet[i.e. the GC stars
exhibit the Na/O anticorrelation]{Carretta2009}. Note that in this
paper only Na abundances are derived for the PAndAS GCs, though O
abundances would also be affected by multiple populations.  However,
the O lines are fairly weak in this wavelength range, and are not
easily detectable in such low S/N spectra.  Lines in other regions
(e.g. the near infrared) would provide more robust integrated O
abundances.

\subsubsection{[Na/Mg]}\label{subsubsec:NaMg}
Na and Mg are expected to form primarily in massive stars
(e.g. \citealt{WoosleyWeaver1995,Timmes1995}); in the absence of
multiple population effects, the abundances of Na and Mg are therefore
expected to correlate.  The MW field stars with $[\rm{Fe/H}] \ga -2$
show a clear trend of increasing [Na/Mg] with
[Fe/H]---\citet{Gehren2006} interpret this as a signature of
increasing Na yields with metallicity.  However, the metal poor ($-2
\la [\rm{Fe/H}] \la -1.2$) Galactic and PAndAS GCs show the opposite
trend: in PA06 and M15 [Na/Mg] is very high, though [Na/Mg] decreases
to subsolar values by $[\rm{Fe/H}] \sim -1.5$.  The GCs therefore
behave in a very different way from the field stars.  Again, this is
likely to be a result of multiple populations, combined with the
metallicity dependency of the Mg/Al anticorrelation
(e.g. \citealt{Carretta2009}).  For these lower mass GCs, at low
metallicities the star-to-star Mg variations are likely to be strong
in these (fairly bright) GCs, driving the integrated [Mg/Fe] down and
[Na/Mg] up.  With increasing [Fe/H], the integrated [Mg/Fe] returns to
its ``primordial'' value, and the [Na/Mg] ratio stabilizes.  The
[Na/Mg] abundances of the PAndAS GCs therefore also suggest that most
of the GCs host multiple populations.  Note that the behaviour of the
massive M31 clusters (from \citealt{Colucci2014}) is similar at the
metal poor end, though their [Na/Mg] ratios are higher at the metal
rich end---again, this may be because the massive, metal rich GCs have
stronger Mg variations.

PA17 remains an unusual object, because its similarly high [Na/Fe] and
[Mg/Fe] ratios place its [Na/Mg] in agreement with the MW field stars,
despite its disagreement in [Mg/Fe].  If PA17's high [Mg/Fe] is caused
by increased supernovae ejecta from massive and/or rapidly rotating
stars, then [Na/Fe] would also be enhanced; thus, PA17 may have an
additional integrated Na enhancement that is caused by contributions
from massive stars.

\begin{figure*}
\begin{center}
\centering
\subfigure{\includegraphics[scale=1.0]{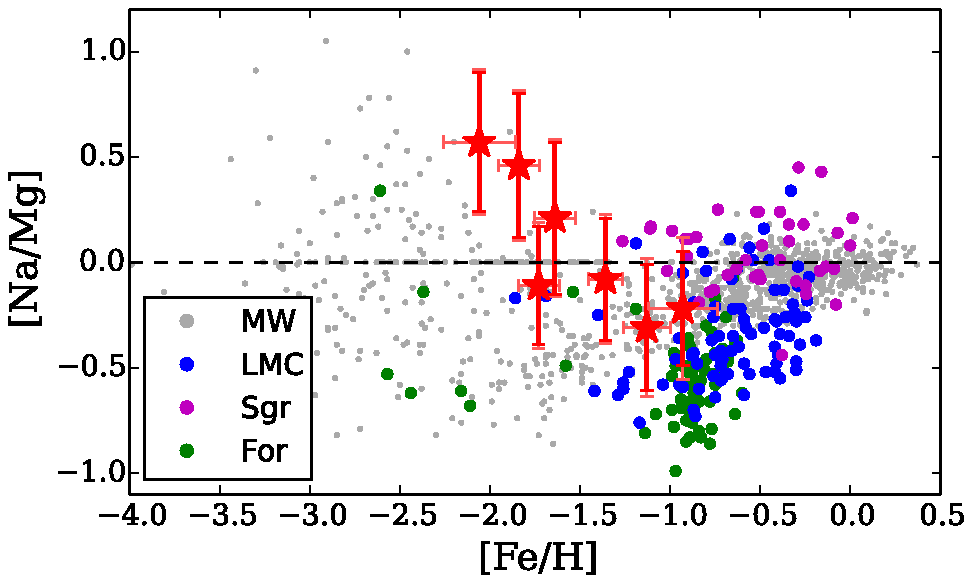}\label{fig:PAndASNaMgFSs}}
\subfigure{\includegraphics[scale=1.0]{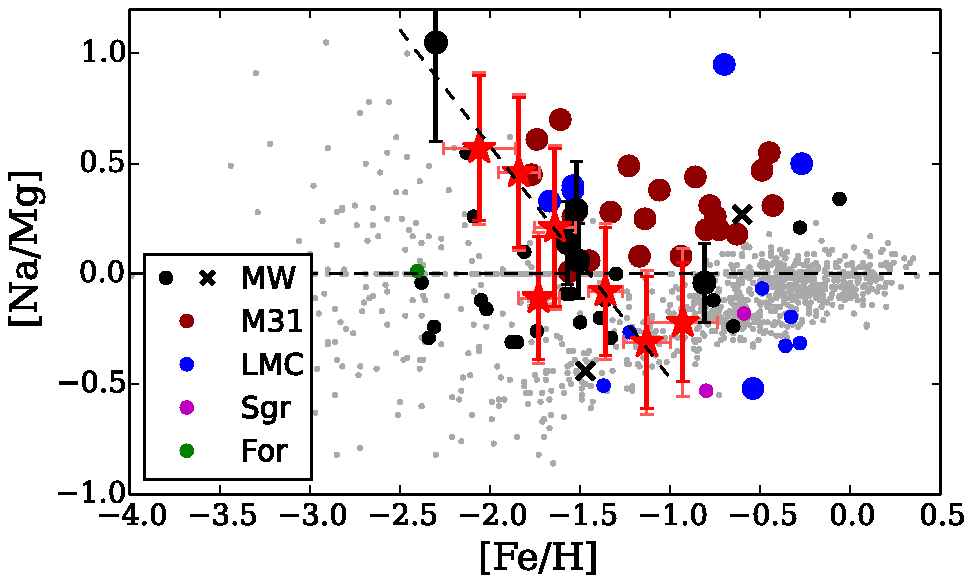}\label{fig:PAndASNaMgGCs}}
\caption{{\it Top: } Comparisons of PAndAS clusters (red stars) to
MW field stars (grey; from the sources in \citealt{Venn2004}, with
supplements from \citealt{Reddy2006}) and dwarf galaxy field stars.
Points and references are as in Figure \ref{fig:PAndASMgFSs}.
{\it Bottom: } Comparisons of PAndAS clusters (red stars) to MW field
stars (grey), MW clusters (black circles), inner halo M31 clusters
(maroon), and dwarf galaxy GCs.  Points and references are as in
Figure \ref{fig:PAndASMgGCs}.  The dashed line shows the clear linear
relationship for the metal poor GCs ($[\rm{Fe/H}] <
-1.2$).\label{fig:PAndASNaMg}}
\end{center}
\end{figure*}

\begin{figure*}
\begin{center}
\centering
\subfigure{\includegraphics[scale=1.0]{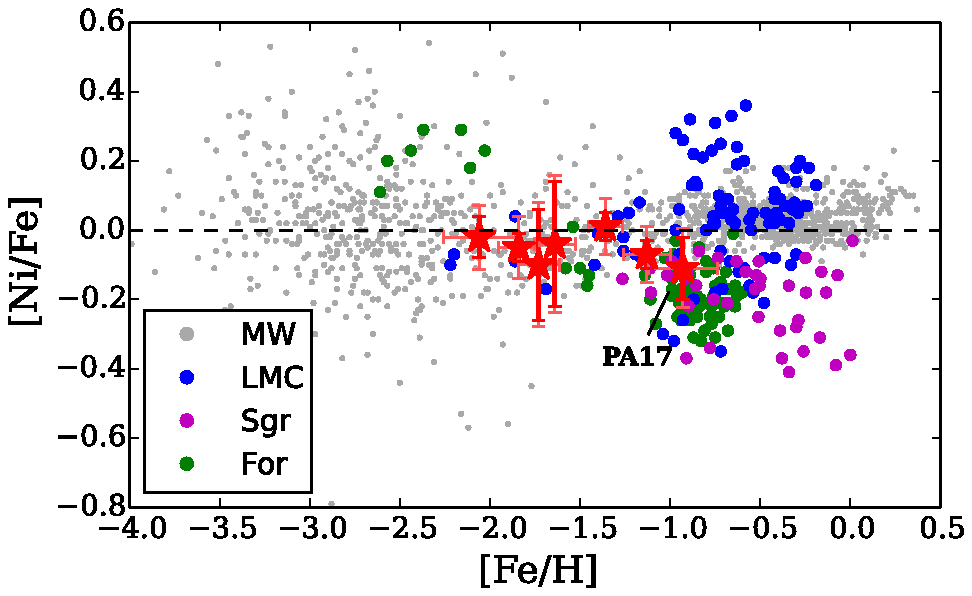}\label{fig:PAndASNi}}
\subfigure{\includegraphics[scale=1.0]{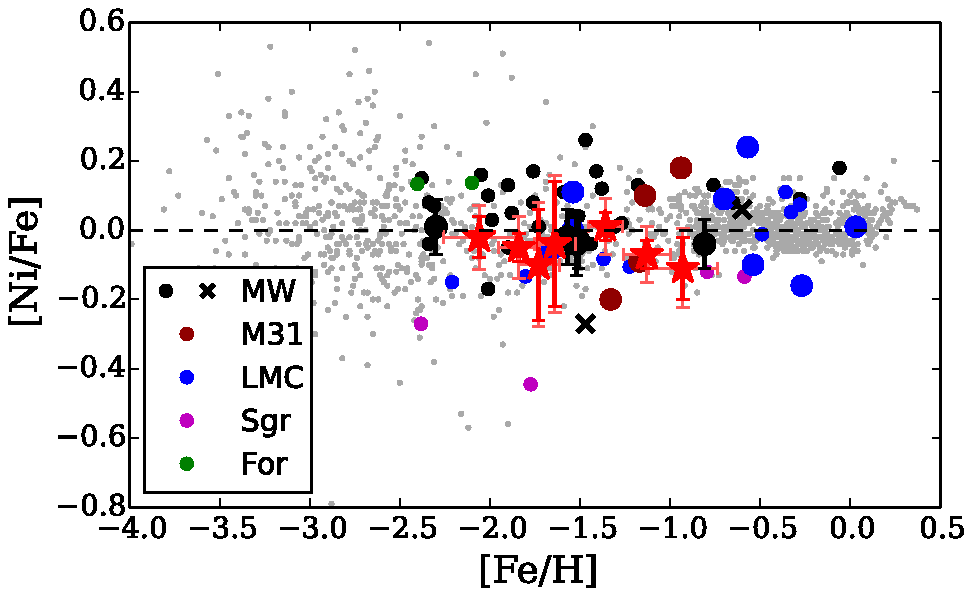}\label{fig:PAndASNiGCs}}
\caption{{\it Top: } Comparisons of Ni abundances in the PAndAS
clusters (red stars) to those in MW field stars (grey), dwarf galaxy
field stars (coloured points), and the IL abundances of the Galactic
clusters (black circles).  Points and references are as in Figure
\ref{fig:PAndASMg}.  {\it Bottom:} Comparisons of [Ni/Fe] ratios in
PAndAS clusters to MW field stars and clusters, Pal~1, and dwarf
galaxy clusters.  Points and references are as in Figure
\ref{fig:PAndASMgGCs}.\label{fig:Ni}}
\end{center}
\end{figure*}

\subsection{Nickel}\label{subsec:PAndASNi}
Nickel is expected to form in core collapse and Type Ia supernovae,
along with Fe \citep{Timmes1995}.  The [Ni/Fe] ratios were determined
with EWs, even though there are few ($\sim 1-7$) \ion{Ni}{1}
lines---however, these lines were carefully checked to ensure that
they were clean (i.e. uncontaminated by cosmic rays, sky lines, or
noise) and were properly measured. The Ni abundances are given in
Table \ref{table:Abunds} and are compared to field stars and other GCs
in Figure \ref{fig:Ni}.  The PAndAS clusters all have [Ni/Fe] ratios
that are consistent with MW and dwarf galaxy stars---however, H23 and
PA17 both have slightly low Ni abundances, in better agreement with
dwarf galaxy stars and GCs. In particular, PA17's
[\ion{Ni}{1}/\ion{Fe}{1}] ratios agree well with the Pal~12 and Ter~7
ratios.

\subsection{Neutron Capture Elements: Ba and Eu}\label{subsec:PAndASNeutronCapture}
\begin{figure*}
\begin{center}
\centering
\includegraphics[scale=1]{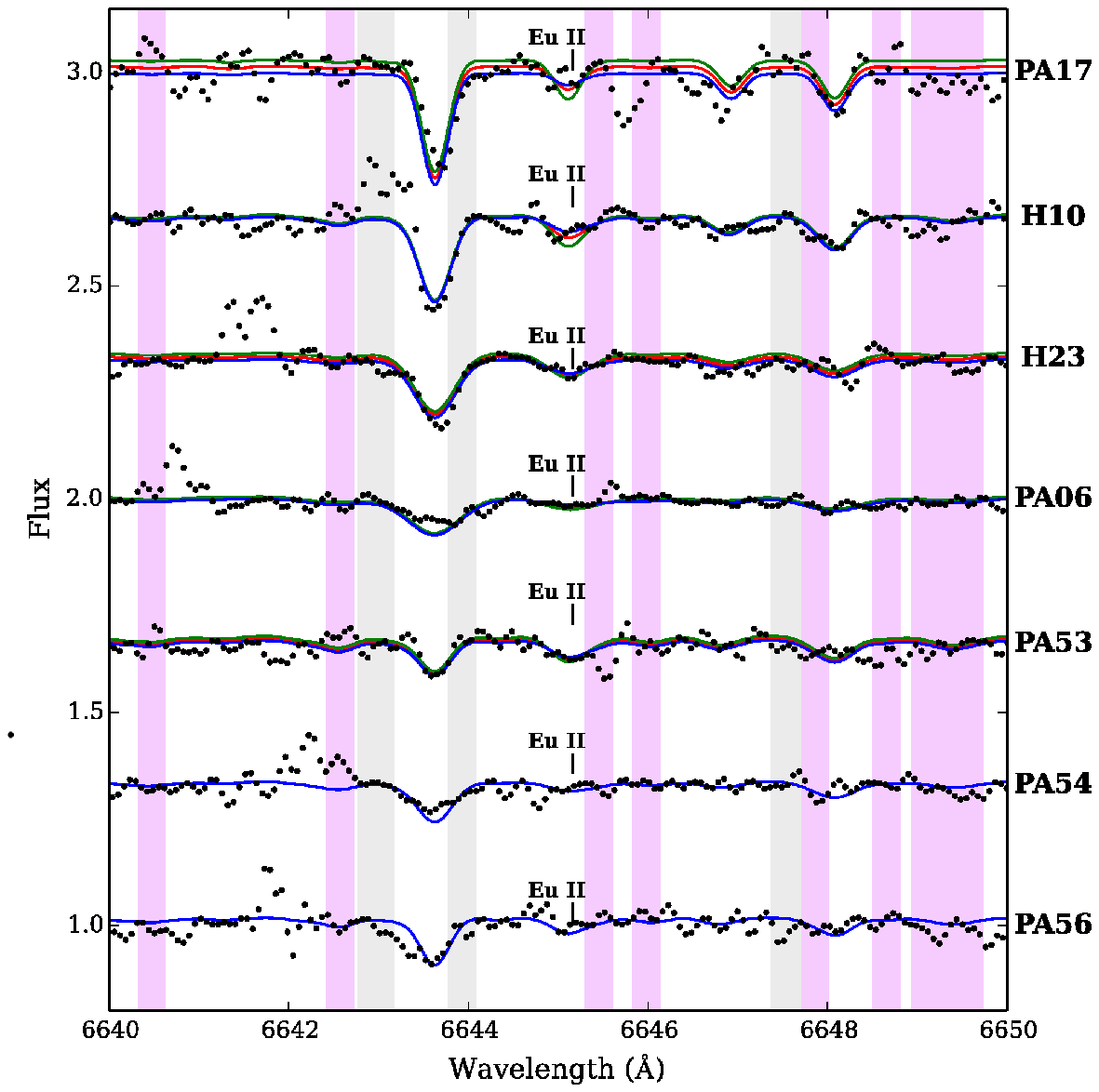}
\caption{Syntheses of the 6645 \AA \hspace{0.025in} \ion{Eu}{2} line.
Grey regions show areas with uncertain HFS, while purple regions
indicate uncertain molecular features.  The red lines show the best
fit, while the green and blue lines show $\pm 1\sigma$ uncertainties.
PA06 and PA56 have only upper limits for the \ion{Eu}{2}
abundance.\label{fig:SynthEu}}
\end{center}
\end{figure*}

The Ba and Eu abundances are determined through spectrum syntheses.
Only two \ion{Ba}{2} lines were considered: the 5853 and 6141~\AA
\hspace{0.025in} lines---the 6496 \AA \hspace{0.025in} line was
removed because of possible NLTE effects which lead to abundance
overestimates \citep{Mashonkina2000}. No molecular lines are
included in these syntheses, since none are identified in those
regions in the Arcturus Atlas.   The \ion{Eu}{2} abundances are
determined through syntheses of the 6645 \AA \hspace{0.025in} line and
include CH and CN molecular lines (see \citealt{Sakari2013}); these
syntheses are shown in Figure \ref{fig:SynthEu}.  HFS and isotopic
components were included for the \ion{Ba}{2} and \ion{Eu}{2} lines, as
described in Section \ref{subsec:DAOSPEC}.

\citet{Sakari2014} demonstrated that the [Ba/Eu] ratio is often more
stable than the individual [Ba/Fe] and [Eu/Fe] ratios to uncertainties
in the underlying population.  Comparisons of [Ba/Eu] ratios 
are shown in Figure \ref{fig:PAndASBaEu}, and illustrate that
metal-rich dwarf galaxy stars have higher [Ba/Eu] ratios than MW field
stars.  PA06 and PA54 only have upper limits on the Eu abundance,
which translates into a lower limit in [Ba/Eu]. The estimated
r-process-only ratio (from \citealt{Burris2000}) is also indicated in
Figure \ref{fig:PAndASBaEu}.  The PAndAS GCs agree with the MW field
stars and GCs, with the exception of PA17 and PA56. PA56 has a low
[Ba/Eu], much like M15.  Both Ba and Eu have been known to vary within
some of the most massive metal poor GCs\footnote{Though note that not
all massive GCs show heavy element variations, e.g. M92
\citep{Cohen2011}.} (such as M15; see, e.g., \citealt{Sneden1997},
\citealt{Roederer2011}); neutron star mergers have been suggested as
the origin of these star-to-star heavy element variations
(e.g. \citealt{TsujimotoShigeyama2014}).  PA56's IL abundances may
also be affected by the presence of heavy element dispersions with the
GCs.   Table \ref{table:Abunds} shows  that PA56 has a moderate Ba
abundance and a high Eu abundance, which leads to a [Ba/Eu] that is
below the r-process only estimate; this indicates that PA56 may host
stars with a significant heavy element dispersion.

PA17's high [Ba/Eu] is in agreement with the Galactic GCs within its
$1\sigma$ errors, though it agrees best with the dwarf galaxy stars
and GCs, Pal~1, and Pal~12.  The high [Ba/Eu] ratios in dwarf galaxies
are typically interpreted as an excess of s-process over r-process
elements.  Its moderately high [Ba/Eu] ratio indicates that PA17 has
received chemical contributions from AGB stars.

[Eu/$\alpha$] is another popular chemical tagging indicator since
dwarf galaxy stars and GCs have higher [Eu/$\alpha$] ratios than MW
stars and GCs at a given [Fe/H]---this has been interpreted as a sign
of an additional r-process site (e.g. \citealt{Letarte2010}) or a
top-light IMF (e.g. \citealt{McW2013}).  Figures \ref{fig:PAndASEuCa}
and \ref{fig:PAndASEuCaGCs} show the [Eu/Ca] ratios (which serve as
[Eu/$\alpha$] indicators) in different environments.  Again, only
upper limits are available for PA06 and PA54.  In general the clusters
are in agreement with the MW and dwarf galaxy field stars and GCs,
with the exception of PA56, which has a high [Eu/Ca], similar to
M15---again, this is likely a signature of star-to-star Eu
variations.

\begin{figure*}
\begin{center}
\centering
\subfigure{\includegraphics[scale=1.0]{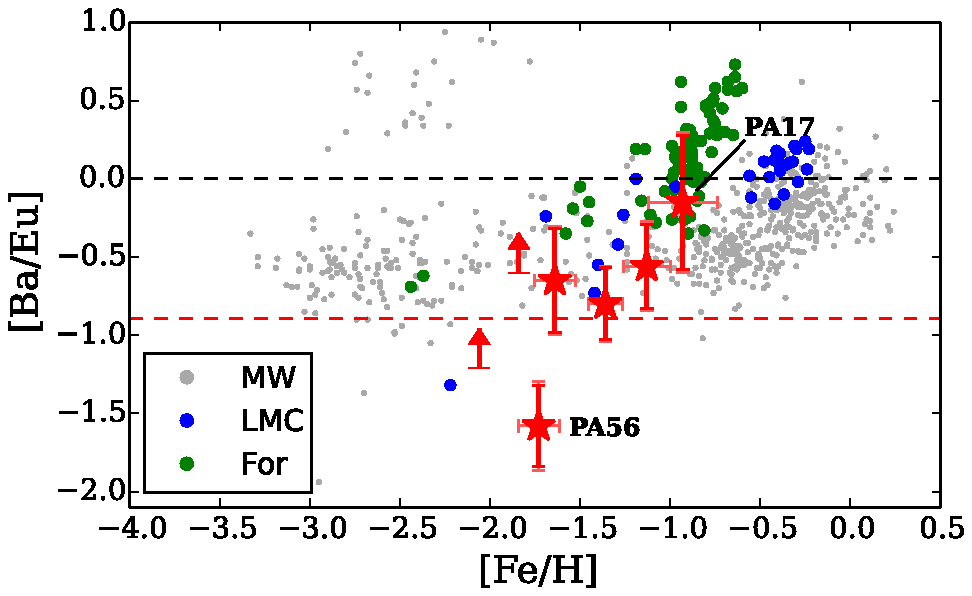}\label{fig:PAndASBaEu}}
\subfigure{\includegraphics[scale=1.0]{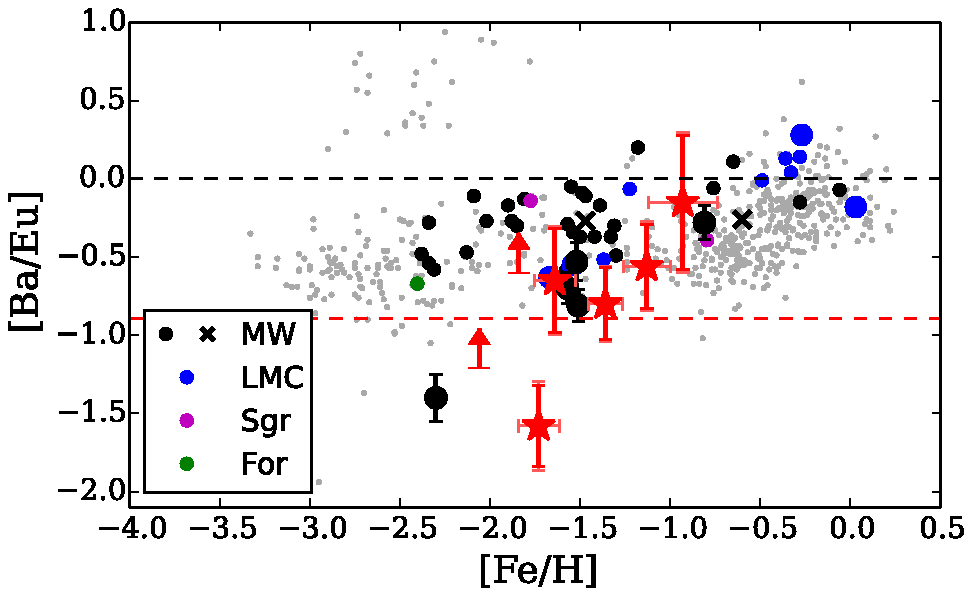}\label{fig:PAndASBaEuGCs}}
\caption{{\it Top: } Comparisons of PAndAS clusters (red stars) to
MW field stars and dwarf galaxy field stars. Points and references are
as in Figure \ref{fig:PAndASMgFSs}.  The dashed red line shows the
r-process-only limit from \citet{Burris2000}.  
{\it Bottom: } Comparisons of PAndAS clusters (red stars) to MW field
stars and GCs from the MW, M31, and various dwarf galaxies.  Points
and references are as in Figure
\ref{fig:PAndASCaGCs}.\label{fig:PAndASBaEu}}
\end{center}
\end{figure*}

\begin{figure*}
\begin{center}
\centering
\subfigure{\includegraphics[scale=1.0]{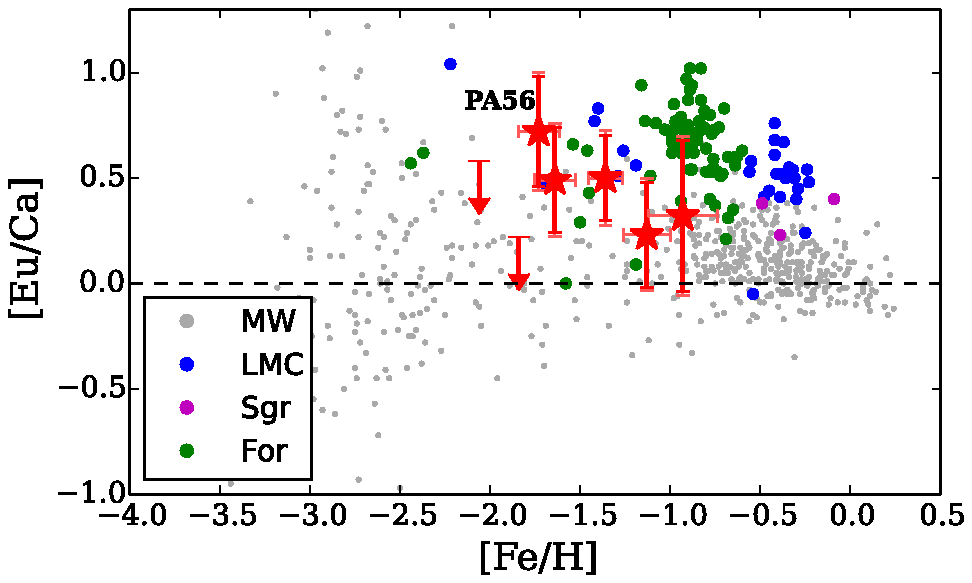}\label{fig:PAndASEuCa}}
\subfigure{\includegraphics[scale=1.0]{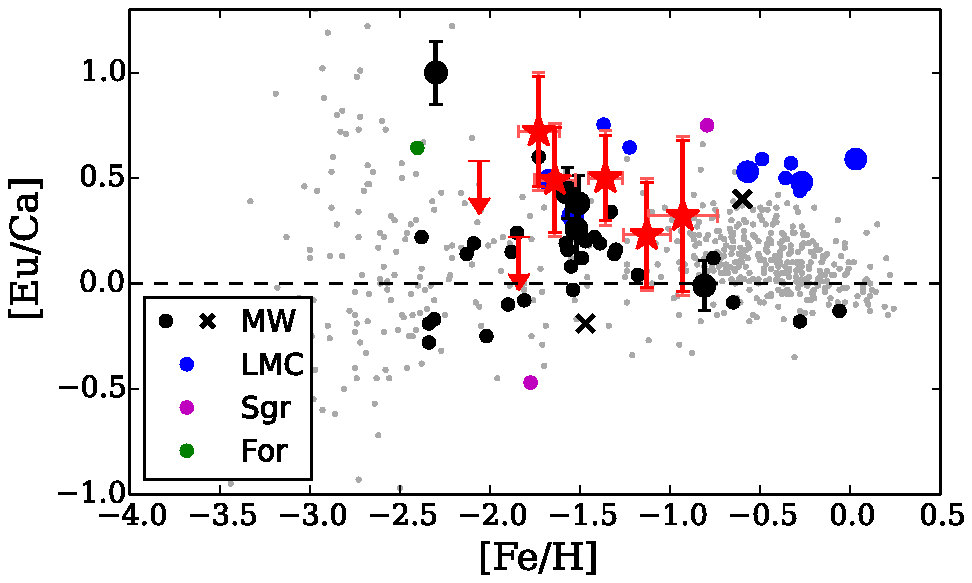}\label{fig:PAndASEuCaGCs}}
\caption{{\it Top: } Comparisons of PAndAS clusters (red stars) to
MW field stars and dwarf galaxy field stars. Points and references are
as in Figure \ref{fig:PAndASMgFSs}.
{\it Bottom: } Comparisons of PAndAS clusters (red stars) to MW field
stars and GCs from the MW, M31, and various dwarf galaxies.  Points
and references are as in Figure
\ref{fig:PAndASCa}.\label{fig:PAndASEuCa}}
\end{center}
\end{figure*}

\section{Discussion}\label{sec:Discussion}
The detailed abundances of stars are dictated by the chemical
composition of the interstellar medium in their host galaxy.  Chemical
comparisons between the PAndAS GCs and the field stars and GCs
associated with other galaxies therefore provide clues about the
nature of a GC's birth environment, such as whether a cluster formed
in a massive galaxy or a dwarf galaxy.  Chemical comparisons can also
determine if the GCs are chemically similar to {\it particular}
galaxies, streams, or other GCs.  If these GCs originated in dwarf
galaxies, the abundances in this paper can therefore be used to infer
the nature of the dwarfs that are currently being accreted into M31's
outer halo.

\subsection{Individual Clusters: A Summary}\label{subsec:PAndASSummary}

\subsubsection{PAndAS Clusters with $[\rm{Fe/H}] > -1.5$}\label{subsec:MRSummary}
The more metal-rich clusters all have higher metallicities than
expected given their large projected distances from the centre of
M31. However, each GC has a unique chemical signature.

\begin{description}
\item[PA17: ] Without {\it HST} photometry, PA17's abundances are more
uncertain than the other GCs; however, the high [Na/Fe] ratio suggests
that PA17 is a ``classical'' GC (under the \citealt{Carretta2009}
definition) with signs of a Na spread (and therefore probably an O
spread as well).  PA17 is the most metal-rich of the target clusters,
at $[\rm{Fe/H}] \sim -0.9$.  Despite the lack of {\it HST} photometry,
systematic errors in [Fe/H] are likely to be $\la 0.2$ dex. PA17's
metallicity alone indicates that it formed in a fairly massive
galaxy---its unusual location far away in the outer halo
($R_{\rm{proj}} \sim 54$ kpc; \citealt{Huxor2014}) suggests that it
formed in a  dwarf galaxy like the LMC or Sgr.\\

PA17's low [Ca/Fe] and high [Mg/Ca] and [Ba/Eu] ratios indicate that
PA17 is more {\it chemically} similar to the LMC stars and clusters
than to those associated with the MW (and presumably to those in M31),
even when systematic offsets are considered.  As discussed earlier,
the [$\alpha$/Fe] and [Ba/Eu] ratios indicate whether a cluster formed
in an environment enriched by Type Ia supernova and AGB star
products.  Combined with its metallicity, PA17's detailed chemical
abundance ratios therefore indicate that it formed in a galaxy which
had sufficient mass to have a fairly high star formation rate or to
populate the high mass end of the IMF.   However, its high [Mg/Fe]
(and possibly its high [Na/Fe]) indicate that PA17 may have been
enriched in ejecta from the most massive stars.  As discussed in
Sections \ref{subsec:PAndASAlpha} and \ref{subsec:PAndASLight} this
suggests that PA17's progenitor galaxy had more massive stars than a
typical dwarf (i.e. it had a top-heavy IMF), was enriched by supernova
ejecta from a rapidly rotating massive star, and/or was uniquely
enriched from the products of a supernova due to inhomogeneous mixing
in the host galaxy.\\

Ultimately, PA17's integrated abundances are chemically distinct from
those of the metal-rich Galactic GC 47~Tuc, and are similar to Pal~1,
the intermediate age LMC clusters, and the accreted Sgr clusters
Pal~12 and Ter~7.\\

\item[H10 and H23: ]  H10 and H23 are the third and second most
metal-rich GCs in the sample of PAndAS GCs, respectively.  Based on
the same arguments as for PA17, this indicates that the clusters
likely originated in LMC- or Sgr-like dwarf galaxies.  Unlike
PA17, H23's chemical abundance ratios are indistinguishable from MW
field stars and clusters and M31 GCs at the same [Fe/H].  H10's
[Ca/Fe] is mildly low while its [Mg/Ca] and [Eu/Ca] ratios are mildly
high---this [Ca/Fe] ratio could indicate that H10 formed in a dwarf
galaxy with a knee in the [Ca/Fe] vs. [Fe/H] relationship near
$[\rm{Fe/H}] \sim -1.5$, similar to LMC intermediate-aged clusters. 
\end{description}

\subsubsection{PAndAS Clusters with $[\rm{Fe/H}] < -1.5$}\label{subsec:MPSummary}
It is more difficult to tag metal-poor GCs ($[{\rm Fe/H}] \la -1.5$)
with chemical abundances, since the chemistries of metal-poor stars in
dwarf galaxies and massive galaxies are not always significantly
different.  The analysis is further complicated by potential
star-to-star Na, Mg, Ba, and Eu variations within the GCs.

\begin{description}
\item[PA53 and PA56: ]  These clusters are fairly close to each other
in projection, and have very similar radial velocities and
metallicities; it is therefore possible that these clusters are
physically associated with one another and were accreted from the same
dwarf.  Their discrepant Mg, Ba, and Eu ratios do not preclude the
possibility that they are related, as the integrated abundances may be
affected by star-to-star variations within the clusters.  Both GCs are
slightly enhanced in Na (with [Na/Fe]$\sim 0.4$), which is likely a
signature of the Na/O anticorrelation.  PA56 also has a very low
[Ba/Eu] (below the r-process only yields), similar to M15, which
supports the idea that it may have star-to-star heavy element
variations.  Both clusters are metal-poor ($[\rm{Fe/H}] \sim -1.7$)
and have [Ca/Fe] ratios slightly lower than the MW field stars,
similar to the Fornax field stars.\\

\item[PA54: ] Although PA54 is extremely close to PA53 in projection,
their discrepant radial velocities imply that these clusters are not
likely to be associated, nor is PA54 likely to be associated with
PA56.  However, PA54 has very similar abundance ratios as PA53 and
PA56, even with only upper limits on [Eu/Fe].  This indicates that
PA54 may have formed in similar conditions.\\

\item[PA06: ]  PA06 is the most metal-poor PAndAS GC in this analysis,
though it is still more metal-rich than the Galactic GC M15.  Again,
PA06's high [Na/Fe] indicates the presence of a Na/O
anticorrelation---additionally, its low [Mg/Fe] hints at a Mg/Al
anticorrelation.  PA06 is similar to the metal-poor GCs in the MW
(particularly M15) and in the dwarf galaxies: it is metal-poor and
$\alpha$-enhanced.  The weakness of its spectral lines means that only
an upper limit can be obtained for \ion{Eu}{2}.
\end{description}

\subsection{Comparisons with M31 Halo Stars}\label{subsec:M31Dwarfs}
The presence of streams in the metal-poor density map of M31's outer
halo (see Figure \ref{fig:PAndASClusterLocations}) implies that some
metal-poor (i.e. low mass) dwarf galaxies are currently being
accreted.  The chemical ratios of PA06, PA53, PA54, and PA56 are all
consistent with an accretion origin in at least one metal-poor dwarf
galaxy.  None of these GCs have been associated with streams based on
their positions.  Given their lack of association with any streams,
if PA06 and PA54 were accreted from dwarf satellites they may be
remnants of ancient accretion events; \citet{Ibata2014} estimate that
$\sim58$\%  of the metal-poor halo ($[\rm{Fe/H}] \la -1.7$) resides in
a ``smooth'' component, which may include these GCs.  

Recall that PA53 and PA56 are located near the Eastern Cloud.  Their
similar kinematics and chemical compositions hint that PA53 and PA56
are associated with each other---however, the numbers of stars around
the clusters (Mackey et al. 2015, {\it in prep.}), their
metallicities, and their detailed chemical abundances indicate that
PA53 and PA56 were also likely associated with the dwarf galaxy that
created the Eastern Cloud.  In this case, the Eastern Cloud must turn
back toward M31, encompassing PA53.  This further suggests that the
Eastern Cloud's progenitor must have been massive enough to form GCs,
though not sufficiently massive to produce a significant metal rich
stellar population.

The presence of the metal-rich Giant Stellar Stream (GSS), the
associated \ion{H}{1} gas, the outer halo stellar mass, and the number
of metal-rich GCs in the outer halo indicate that M31 likely
experienced a minor merger with a Sgr or LMC-mass galaxy
(e.g. \citealt{Fardal2013}, \citealt{Lewis2013}, \citealt{Bate2014}).
\citet{Ibata2014} find that 86\% of the most metal-rich stars
($[\rm{Fe/H}] \ga -0.5$) in the outer halo are associated with
substructure from an accreted companion.  Though this number decreases
with metallicity, the fraction of accreted stars still remains high:
78\% of the stars in PA17's metallicity bin ($-1.1 \la [\rm{Fe/H}] \la
-0.5$) are expected to be associated with coherent, accreted
structures, while 58\% of the stars in H10 and H23's metallicity bin
($-1.7 \la [\rm{Fe/H}] \la -1.1$) are expected to be in streams.  It
is therefore likely that these comparatively metal rich GCs were
associated with dwarf satellites.  H23 has been tentatively linked to
a stream near the GSS (Stream D; see \citealt{Veljanoski2014}) based
on its position, though its kinematics indicate that such an
association is unlikely; neither H10 nor PA17 have been linked to any
visible stellar streams, though {\it HST} imaging of fields around H10
and H23 reveal extremely metal-rich populations of field stars,
suggesting that the GCs may be located on low surface brightness
streams \citep{Richardson2009}.

H10 is located near the SW Cloud, an overdensity of stars to the
southwest of M31.\footnote{Note that though PA17's proximity to PA14
hints at an association with the SW Cloud, its drastically different
radial velocity ($-260$ km s$^{-1}$) from PA14 ($-363$ km s$^{-1}$;
\citealt{Veljanoski2014}) and high [Fe/H] ($-0.93$ compared to
$-1.30$; \citealt{Mackey2013}) make it unlikely that PA17 was
associated with the SW Cloud progenitor.}  With photometry,
\citet{Bate2014} estimate that the cloud hosts a metal-rich population
($[\rm{Fe/H}] \sim -1.3$), and that the progenitor galaxy was a fairly
bright dwarf ($M_V \sim -12$, which is slightly fainter than For and
Sgr).  Three GCs (PA7, PA8, and PA14) appear to be kinematically
associated with each other and with the SW Cloud
\citep{Mackey2013,Mackey2014,Veljanoski2014,Bate2014}.    The SW Cloud
does extend to the southeast, and H10 lies at the end of this
extension (see Figure 2 in \citealt{Bate2014}).  H10's radial velocity
($-352$ km s$^{-1}$) agrees with the stream's radial velocity and that
of the GCs, and even agrees with the velocity gradient noted by Bate
et al. (where the northernmost GC, PA7, is moving toward the MW faster
than PA8 and PA17).  Furthermore, H10's $[\rm{Fe/H}] \sim -1.4$
agrees very well with the Cloud and its GCs, and its detailed
abundances support the idea that it could have originated in a dwarf
galaxy with the mass of For or Sgr.  Thus, it is likely that H10 is
associated with PA7, PA8, PA14, and the SW Cloud.

As mentioned above, H23 has been spatially associated with Stream D
based on its position, though its velocity disagrees with the other
GCs in the stream.  Its metallicity, $[\rm{Fe/H}] = -1.1$, now further
suggests that H23 is not likely to be associated with that stream,
since Stream D is primarily composed of metal-poor stars with
$[\rm{Fe/H}] < -1.1$ \citep{Ibata2014}.  It is more likely that H23
came from the progenitor of the GSS, along with PA17.
\citet{Fardal2013} perform N-body simulations to reproduce the GSS
and other stellar debris. These simulations indicate that a massive
progenitor was accreted during multiple orbits, culminating in a
recent final accretion ($\sim 760$ Myr ago).  Many of the metal-rich
PAndAS GCs might have been stripped early on, and may no longer be
associated with any bright, coherent substructure.  Given their
metallicities and other chemical abundances, it is possible that H23
and PA17 both formed in the GSS progenitor. If H23 and PA17 were born
in the same dwarf galaxy, their [Ca/Fe] ratios clearly imply a
``knee'' at $[\rm{Fe/H}] \sim -1.1$ to -1.3, similar to the LMC stars
and clusters.

\section{Conclusions}\label{sec:Conclusions}
This paper has presented integrated Fe, Na, Mg, Ca, Ti, Ni, Ba, and Eu
abundances of seven outer halo M31 GCs, five of which were discovered
in the Pan-Andromeda Archaeological Survey. These detailed IL chemical
analyses of PAndAS clusters have identified GCs in an extragalactic
system that may have been accreted from dwarf galaxies.  Detailed
investigations of the chemical abundance ratios of individual targets
such as this are only possible with {\it high resolution}
spectroscopy.

The PAndAS cluster abundances suggest that these outer halo M31 GCs
may have been accreted from multiple dwarf galaxies:
\begin{itemize}
\item The metal-rich GC PA17 is chemically more similar to the LMC
stars and clusters than to MW field stars and clusters, suggesting
that it originated in an LMC-like dwarf galaxy.  H23's abundances are
indistinguishable from MW field stars and clusters, though its
metallicity and location suggest it may have originated in a massive
dwarf galaxy.  PA17 and H23 may have been accreted along with the
progenitor of the metal-rich Giant Stellar Stream.\\

\item The intermediate metallicity GC H10 has a location, metallicity,
and radial velocity that agree well with the SW Cloud and its GCs,
which may have originated in a Sagittarius or Fornax sized progenitor
\citep{Bate2014}. H10's chemical abundance ratios support the
suggestion that it may have formed in a dwarf galaxy.\\

\item PA53, PA54, and PA56 have abundances and metallicities that are
more typical of an intermediate mass dwarf galaxy like For.  This
suggests that they are currently being accreted from at least one
metal-poor dwarf galaxy and could be associated with one or more of
the coherent, metal-poor streams observed in PAndAS.  Based on their
chemistries and radial velocities PA53 and PA56 could be associated
and are likely to have been accreted from the dwarf galaxy that
created the Eastern Cloud.  Despite its proximity in projection, it is
unlikely that PA54 is associated with either PA53 or PA56.\\

\item PA06's metallicity makes it a difficult target for chemical
tagging analyses, since the chemistries between dwarf and massive
galaxies have likely not had sufficient time to diverge at
$[\rm{Fe/H}] \sim -2$.  Several of its integrated abundances do not
agree with MW or dwarf galaxy stars, suggesting that strong
star-to-star chemical variations are present in the cluster.\\ 
\end{itemize}
\noindent Thus, this chemical tagging analysis is consistent with the
observation that M31's outer halo GC system is being built up by
accretion of dwarf satellites.  In addition, this detailed abundance
analysis provides additional information on the nature of these
progenitor systems that is, at present, very difficult to obtain from
the field populations.

\section*{Acknowledgments}
The authors thank Don VandenBerg for his suggestion to investigate the
RGB slope calibration, for providing the isochrones in {\it HST}
magnitudes, and for all his comments.  The authors also thank Judy
Cohen for providing a helpful referee report that contributed to this
analysis and George Wallerstein for reading the manuscript.  CMS
acknowledges funding from the Natural Sciences \& Engineering Research
Council (NSERC), Canada, via the Vanier CGS program.  KAV acknowledges
funding through the NSERC Discovery Grants program.   ADM is grateful
for support by an Australian Research Fellowship (Grant DP1093431)
from the Australian Research Council.  The Hobby-Eberly Telescope
(HET) is a joint project of the University of Texas at Austin, the
Pennsylvania State University, Stanford University,
Ludwig-Maximilians-Universit\"{a}t M\"{u}nchen, and
Georg-August-Universit\"{a}t G\"{o}ttingen. The HET is named in honour
of its principal benefactors, William P. Hobby and Robert
E. Eberly. The authors wish to thank the night operations staff of the
HET for their assistance and expertise with these unusual
observations.  This work has made use of BaSTI web tools.

\footnotesize{

}

\clearpage
\normalsize
\appendix

\section{RGB Slope vs. Metallicity: A Calibration with Galactic
Globular Cluster Optical CMDs}\label{appendix:RGBslopes}
\begin{figure*}
\begin{center}
\centering
\includegraphics[scale=1.2]{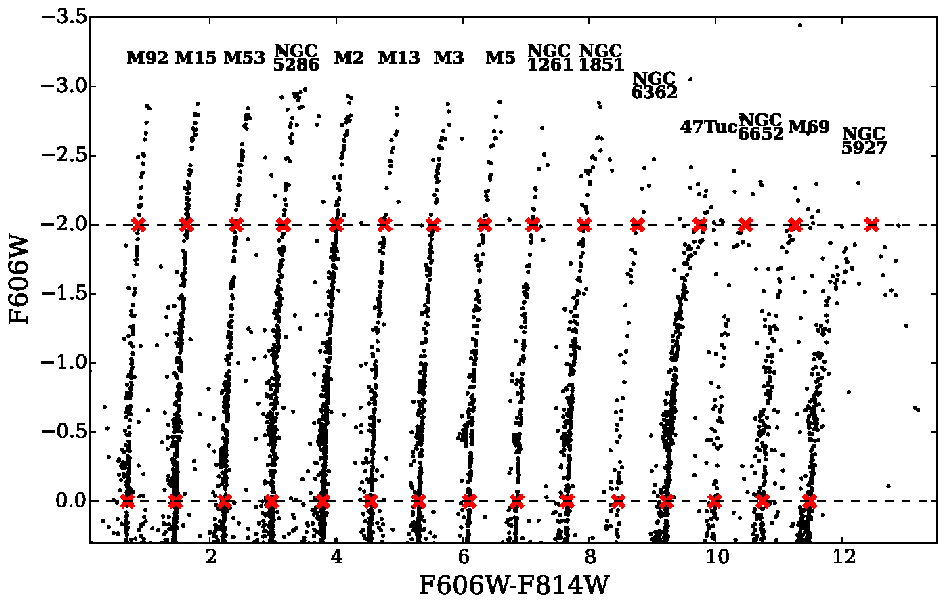}
\caption{F606W, F814W photometry of Galactic GC RGBs, from the ACS
Globular Cluster Survey
\citep{Sarajedini2007,Anderson2008}. The cluster HBs were aligned, and
the colours at two magnitudes ($M_V = 0$ and $M_V = -2$, shown as the
horizontal dashed lines) were determined for each cluster; these
values are shown as red crosses.  The clusters are offset in the plot,
and are ordered by metallicity (from
\citealt{Carretta2009}).\label{fig:AllRGBs}}
\end{center}
\end{figure*}

It is well established that the slope of a cluster's RGB is correlated
with cluster metallicity (e.g. \citealt{Sarajedini1994}).  This
appendix discusses the calibration of RGB slope vs. [Fe/H], utilizing
high quality {\it HST} CMDs of Galactic GCs from the ACS Survey of
Galactic Globular Clusters
\citep{Sarajedini2007,Anderson2008}.\footnote{\url{http://www.astro.ufl.edu/~ata/public_hstgc/}}
For this calibration 15 Galactic GCs with high quality CMDs and
well-populated RGBs that span a wide range in [Fe/H] were selected;
priority was given to clusters with low foreground reddening.  Using
the original F606W and F814W {\it HST} magnitudes, the distance moduli
from \citet{Vandenberg2013} were used to overplot clusters, and
distance moduli and reddening values were adjusted so that cluster HBs
overlapped.  The average RGB colours were then determined at two
magnitudes ($M_V = 0$ and $M_V = -2$), and RGB slopes were
calculated. These points are shown in Figure \ref{fig:AllRGBs}, while
the slopes are listed in Table \ref{table:RGBslopes}.

Uncertainties in RGB slope were estimated based on the uncertainties
in colour at the two magnitudes.  The metal-poor clusters have small
uncertainties in RGB colour, though this translates into large
uncertainties in slope because the RGBs are steeper.  The metal-rich
clusters have very uncertain RGB colours, but because the RGBs are
flatter the uncertainty in slope is much lower.  Note that at higher
[Fe/H] ($\ga -0.7$) a linear fit to the RGB is no longer a good
approximation to the actual shape of the RGB---thus, this calibration
is likely to break down at the metal rich end.

\begin{figure*}
\begin{center}
\centering
\includegraphics[scale=0.6]{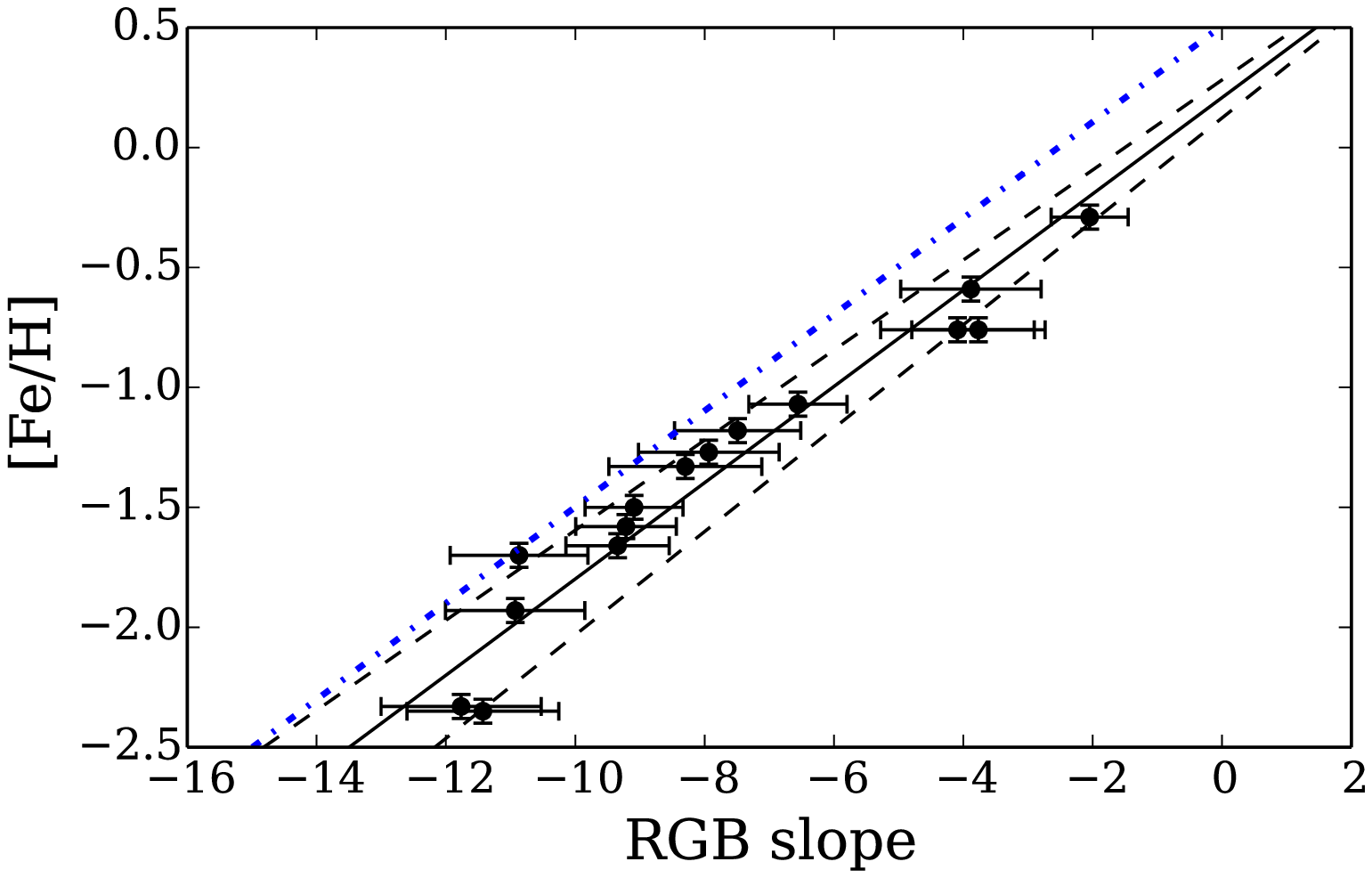}
\caption{RGB slope versus [Fe/H] for the Galactic GCs.  Slopes are
determined using the points in Figure \ref{fig:AllRGBs}; slope errors
were estimated based on the uncertainties in these RGB colours.
\label{fig:MWRGBslopes}}
\end{center}
\end{figure*}

\begin{figure*}
\begin{center}
\centering
\subfigure[]{\includegraphics[scale=0.5]{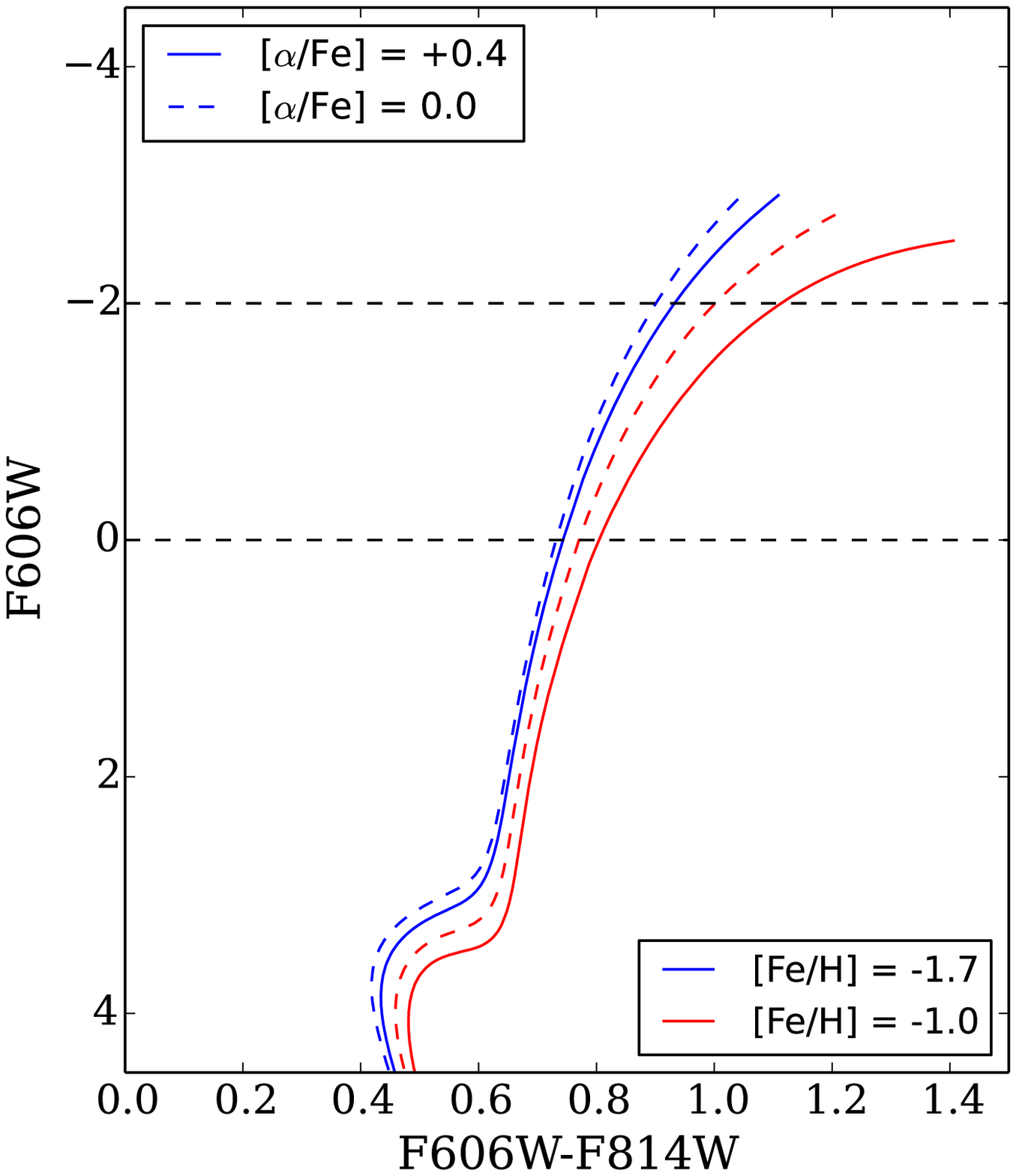}\label{fig:AlphaIsosA}}
\subfigure[]{\includegraphics[scale=0.5]{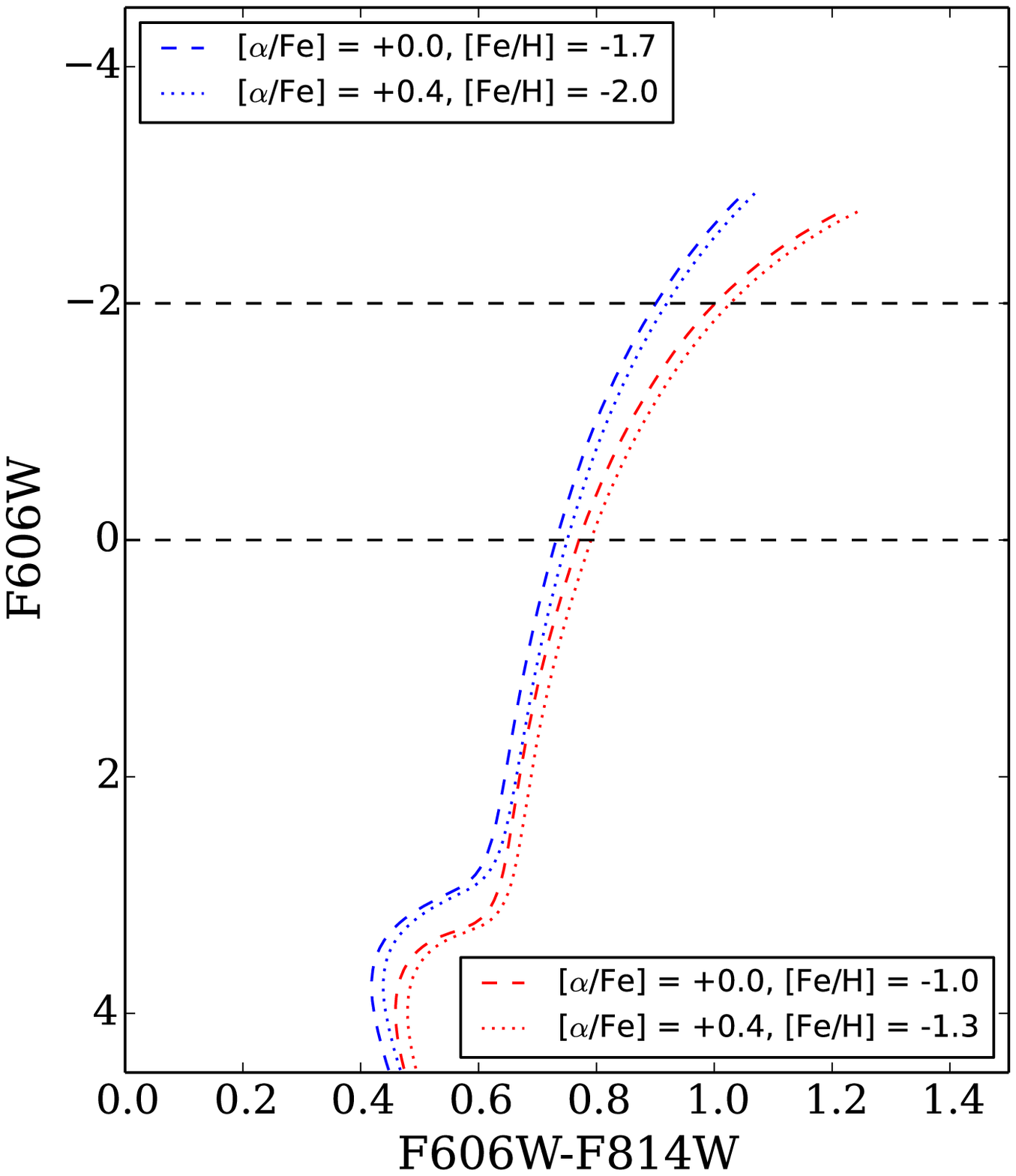}\label{fig:AlphaIsosB}}
\caption{Victoria-Regina F606W, F814W isochrones from
\citet{Vandenberg2014}.  {\it Left: } Metal poor ($[\rm{Fe/H}] =
-1.7$, in blue) and metal rich ($[\rm{Fe/H}] = -1.0$, in red)
isochrones at two different values of [$\alpha$/Fe].  The solid lines
show $[\alpha/\rm{Fe}] = +0.4$, while the dashed lines show
$[\alpha/\rm{Fe}] = 0.0$.  Isochrones with low [$\alpha$/Fe] have
steeper RGB slopes at a fixed metallicity.  {\it Right: } Isochrones
illustrating that the steeper slopes from $[\alpha/\rm{Fe}] = 0.0$
isochrones can be reproduced by lowering [Fe/H] by 0.3 dex while
maintaining $[\alpha/\rm{Fe}] = +0.4$.  The dashed lines show the same
isochrones as in Figure \ref{fig:AlphaIsosA}, while the dotted lines
show isochrones with $[\alpha/\rm{Fe}] = +0.4$ and $\Delta[\rm{Fe/H}]
= -0.3$ dex.  The dotted isochrones are offset by $+0.02$ magnitudes
so that both isochrones can be seen.  This shows that an [Fe/H] offset
of -0.3 dex can be used in the RGB slope calibration to approximate
the effects of lower [$\alpha$/Fe].
\label{fig:AlphaIsos}}
\end{center}
\end{figure*}

For each cluster the [Fe/H] ratios from \citet{CarrettaFe} were
adopted; [Fe/H] uncertainties of 0.05 dex were assumed.  Figure
\ref{fig:MWRGBslopes} shows the relationship between RGB slope and
cluster [Fe/H].  The linear least squares fit to the points is shown
with the solid line, while the dashed lines show the uncertainty in
the fit.  The trend is clear: metal-poor GCs have steeper RGBs than
more metal-rich GCs.  Non-standard chemical abundance mixtures (in,
e.g., C, N, O or $\alpha$-elements like Mg or Si) can affect the shape
of the RGB (e.g. \citealt{Salaris1993,Vandenberg2012}), which would
affect where a cluster falls in the plot.  The adopted [$\alpha$/Fe]
ratio is particularly important: the qualitative effects of
[$\alpha$/Fe] on the F606W, F814W RGB slope are illustrated in Figure
\ref{fig:AlphaIsos} using the Victoria-Regina isochrones from
\citet{Vandenberg2014} and the colour transformations of
\citet{CasagrandeVandenberg2014}.  Figure \ref{fig:AlphaIsosA} shows
that at a given [Fe/H], clusters with $[\alpha/\rm{Fe}] = 0$ have
steeper RGBs than clusters with $[\alpha/\rm{Fe}] = +0.4$; this offset
is greater at higher metallicity.  Furthermore, Figure
\ref{fig:AlphaIsosB} illustrates that the slope difference can be
mimicked by lowering [Fe/H] by 0.3 dex while maintaining
$[\alpha/\rm{Fe}] = +0.4$ (note that in Figure \ref{fig:AlphaIsosB}
the dotted isochrones have been shifted by 0.02 magnitudes so that
both the $\alpha$-enhanced and $\alpha$-normal isochrones can be
seen).  The blue dot-dashed line in Figure \ref{fig:MWRGBslopes} shows
the effects of lowering [$\alpha$/Fe] by 0.4 dex, which is
approximated by lowering the [Fe/H] by 0.3 dex (also see
\citealt{Salaris1993}).

\begin{figure*}
\begin{center}
\centering
\includegraphics[scale=0.5]{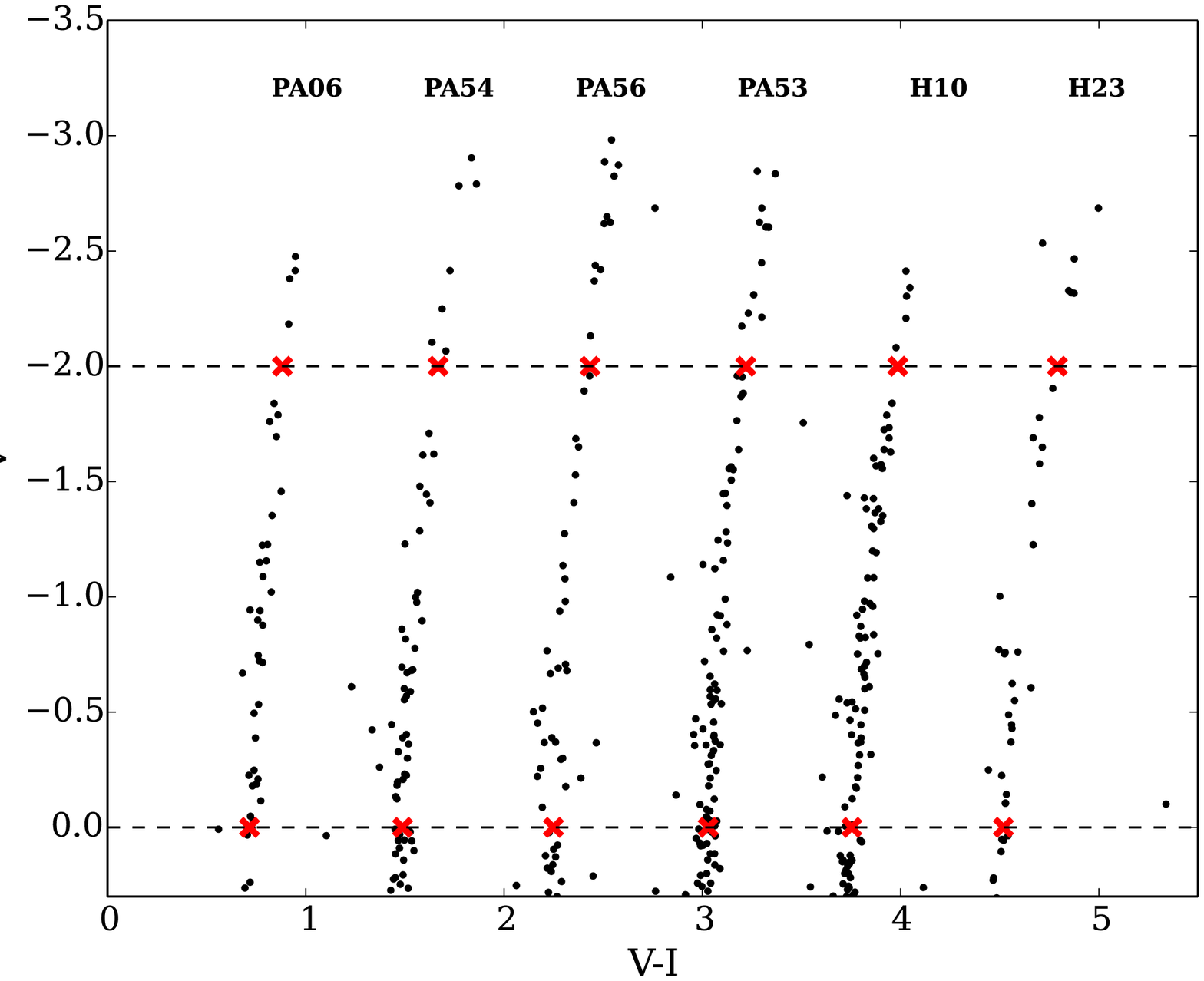}
\caption{F606W, F814W {\it HST} photometry of PAndAS GCs, from Mackey
et al. ({\it in prep.}). The cluster HBs were aligned, and
the colours at two magnitudes ($M_V = 0$ and $M_V = -2$, shown as the
horizontal dashed lines) were determined for each cluster; these
values are shown as red crosses.  The clusters are offset in the plot,
and are ordered by metallicity.\label{fig:PAndASRGBs}}
\end{center}
\end{figure*}

The RGBs of the PAndAS clusters are shown in Figure
\ref{fig:PAndASRGBs}, while the RGB slopes and spectroscopic
metallicities are shown in Figure \ref{fig:RGBslopes}.  The
photometric uncertainties, sparseness of the CMDs, and difficulties in
determining distance moduli (particularly for PA56) make the RGB
slopes more uncertain for the PAndAS clusters, compared to the
Galactic GCs.  However, all PAndAS GCs agree with the MW RGB slope
relation in Figure \ref{fig:MWRGBslopes} within their errors.  This
relationship illustrates that the spectroscopic metallicities are
consistent with the observed RGB slopes in the {\it HST} CMDs.  All of
the clusters in Figure \ref{fig:RGBslopes} lie above the MW relation
closer to the low [$\alpha$/Fe] value, hinting at the possibility of
low [$\alpha$/Fe] ratios in the PAndAS targets.

\begin{table}
\centering
\begin{center}
\caption{Galactic GC RGB slopes.\label{table:RGBslopes}}
 \newcolumntype{d}[1]{D{,}{\pm}{#1} }
  \begin{tabular}{@{}lcccd{5}@{}}
  \hline
Cluster & & [Fe/H] & & \multicolumn{1}{c}{\phantom{----}RGB slope} \\
\hline
M92      & & $-2.35\pm 0.05$ & & -11.429,1.172 \\ 
M15	 & & $-2.33\pm 0.05$ & & -11.765,1.238  \\ 
M53	 & & $-1.93\pm 0.05$ & & -10.929,1.077  \\ 
NGC 5286 & & $-1.70\pm 0.05$ & & -10.870,1.066  \\ 
M2	 & & $-1.66\pm 0.05$ & & -9.346 ,0.799 \\ 
M13	 & & $-1.58\pm 0.05$ & & -9.217 ,0.778 \\ 
M3	 & & $-1.50\pm 0.05$ & & -9.091 ,0.758 \\ 
M5	 & & $-1.33\pm 0.05$ & & -8.299 ,1.181 \\ 
NGC 1261 & & $-1.27\pm 0.05$ & & -7.937 ,1.087 \\ 
NGC 1851 & & $-1.18\pm 0.05$ & & -7.491 ,0.976 \\ 
NGC 6362 & & $-1.07\pm 0.05$ & & -6.557 ,0.760 \\ 
47 Tuc	 & & $-0.76\pm 0.05$ & & -3.766 ,1.031 \\ 
NGC 6652 & & $-0.76\pm 0.05$ & & -4.090 ,1.187 \\ 
M69	 & & $-0.59\pm 0.05$ & & -3.883 ,1.086 \\ 
NGC 5927 & & $-0.29\pm 0.05$ & & -7.491 ,0.595 \\ 
\hline
\end{tabular}\\
\end{center}
\medskip
\raggedright {\bf References: } Cluster [Fe/H] ratios are from
\citet{CarrettaFe}.\\
\end{table}

\begin{table}
\centering
\begin{center}
\caption{PAndAS GC RGB slopes.\label{table:PAndASslopes}}
 \newcolumntype{d}[1]{D{,}{\pm}{#1} }
  \begin{tabular}{@{}lcccd{5}@{}}
  \hline
Cluster & & [Fe/H] & & \multicolumn{1}{c}{\phantom{----}RGB slope} \\
\hline
PA06      & & $-2.06\pm 0.10$ & & -12.0482,2.658 \\ 
PA54      & & $-1.84\pm 0.10$ & & -11.173,1.405 \\ 
PA56      & & $-1.73\pm 0.10$ & & -10.811,2.982 \\ 
PA53      & & $-1.64\pm 0.10$ & & -10.526,1.238 \\ 
H10       & & $-1.40\pm 0.10$ & &  -8.621,0.813 \\ 
H23       & & $-1.12\pm 0.10$ & & -7.353 ,1.268 \\ 
\hline
\end{tabular}\\
\end{center}
\medskip
\raggedright {\bf References: } Cluster [Fe/H] ratios are from
the [\ion{Fe}{1}/H] abundances in Table \ref{table:Abunds}.\\
\end{table}

\end{document}